\documentclass[a4paper,12pt]{article}
\pdfoutput=1 

\usepackage{jheppub} 

\usepackage[T1]{fontenc}
\usepackage[italian,english]{babel}
\usepackage{hyperref}
\usepackage{ifpdf}
\usepackage{subfigure}
\usepackage{amssymb}
\usepackage{amsfonts}
\usepackage{epsf}
\usepackage{rotating}
\usepackage{graphicx}
\usepackage{amsmath}
\usepackage{fancyhdr}
\usepackage{lineno}
\usepackage{dutchcal}
\usepackage{babel}
\usepackage{graphics}
\usepackage{pstricks}
\usepackage{color}
\usepackage{multirow}

\newcommand{\nn}{\nonumber}
\newcommand{\lsim}{\mathrel{\mathop{\kern 0pt \rlap
  {\raise.2ex\hbox{$<$}}}
  \lower.9ex\hbox{\kern-.190em $\sim$}}}
\newcommand{\gsim}{\mathrel{\mathop{\kern 0pt \rlap
  {\raise.2ex\hbox{$>$}}}
  \lower.9ex\hbox{\kern-.190em $\sim$}}}

\newcommand{\be}{\begin{equation}}
\newcommand{\ee}{\end{equation}}
\newcommand{\bea}{\begin{eqnarray}}
\newcommand{\eea}{\end{eqnarray}}

\title{Two-Loop Renormalization Group Equations and Gauge Couplings Unification}

\author[a]{Antonio Costantini}

\affiliation[a]{Dipartimento di Matematica e Fisica "Ennio De Giorgi", Universit\`a del Salento and INFN-Lecce, Via Arnesano, 73100 Lecce, Italy}

\emailAdd{antonio.costantini@le.infn.it}

\abstract{We analize the impact of two-loop renormalization group equations on the $SU(3)_c\times SU(2)_w\times U(1)_Y$ gauge couplings unification in various supersymmetric theories. In general the presence of superfields in higher representation than the doublet spoil the gauge couplings unification at one-loop. The situation is more interesting when the renormalization group equations are calculated at two-loop. In this case we show that the unification of the gauge couplings can be achieved for models with triplet superfield(s). In the analysis of the models with triplet superfield(s) we show that the dimensionless couplings do not have a Landau pole in their evolution at high energies but they run to a nontrivial ultraviolet fixed point.}

\begin{document}
\maketitle
\flushbottom

\section{Introduction}

Grand Unified Theories (GUTs) are an appealing class of theories which exhibit, among other features, the unification of the gauge coupling of the Standard Model (SM) gauge group $SU(3)_c\times SU(2)_w\times U(1)_Y$. The unification is suggested by the approximate convergence of the gauge couplings in the SM at very-high energies and it is more accurate in the context of the Minimal Supersymmetric Standard Model (MSSM). It remains one of the most important reasons to consider supersymmetric extensions of the SM. In GUTs the gauge group of the SM is a subgroup of a larger group and, among the others, the most common studied are $SO(10)$, $SU(6)$, $SU(5)$, $E_6$, $SU(3)_c\times SU(2)_L \times SU(2)_R \times U(1)_{B-L}$. The most simple one is $SU(5)$ and the known SM particles fit into the {\bf 24} and the ${\bf \bar 5}+{\bf 10}+{\bf 1}$ representations.

In this paper we will discuss the impact of the two-loop renormalization group (RG) equations on the unification of the gauge couplings of $SU(3)_c\times SU(2)_w\times U(1)_Y$.  The RG equations at two-loop order for a general supersymmetric model are well known in the literature \cite{martinRG}. They are related to the matter content of the model itself and in particular to the representation to which each chiral superfield belong. We will focus our attention on the running of the dimensionless parameters including the gauge couplings, the Yukawa couplings of the known quarks and the dimensionless couplings of the scalar superfields, which we will shortly call $\vec \lambda$. The evolution of the gauge couplings can depend crucially on $\vec \lambda$ because the two-loop RG equations for the dimensionless parameters of any model represent a set of coupled differential equations. The mutual interdependence can affect drastically the running of the dimensionless couplings in the so-called $\lambda$SUSY models, where the electroweak values of the dimensionless couplings of the superpotential are $\vec \lambda\gsim1$. 

In $\lambda$SUSY models the running of the dimensionless parameters, usually calculated at one-loop, present the appearance of a Landau pole well below the Planck scale. The two-loop running can behave very differently and in this scenario there will be no Landau pole in the evolution of the couplings, which can remain perturbative until the Planck scale.

We will show that the dimensionless couplings, and in particular $\vec \lambda$, can run to a ultraviolet (UV) fixed point. This kind of evolution for $\vec \lambda$ can affect the running of the gauge couplings, leading to the gauge couplings unification for models that do not have it if the RG equations are calculated at one-loop. The presence of UV fixed point in the evolution of the parameters is not a new idea. It has been considered in the context of supersymmetric gauge theories for the gauge couplings where the analysis has been done at higher loop level because at one- and two-loop order the supersymmetric gauge theories do not exhibit such behaviour \cite{martinUVfixed}. Recently the hypothesis of UV fixed point has been taken into account in extensions of the SM with large number of fermions in order to prove the asymptotic safety of these extensions \cite{sannino0,sannino1,sannino2,sanninostrumia}. However our analysis will focus on the connection between the presence of UV fixed point in the evolution of $\vec\lambda$ and the unification of the gauge couplings when superfield(s) in higher representation of $SU(2)_w$, in particular triplet(s), are present. Both the proprieties arise if the RG equations are considered at two-loop order.

The paper is organized as follows. In section \ref{secmssm} we will briefly present the results of the evolution of the gauge couplings in the MSSM and in section \ref{secnmssm} we will show how the RG equations calculated at two-loop can affect the running of the dimensionless parameters in the context of the NMSSM. In section \ref{sectriplet} we will present the main result of the paper, \textit{i.e.} the gauge couplings unification at two-loop order in models where the unification is not present at one-loop, such as models with triplet(s) superfields. In section \ref{sectnssm} we will discuss the case where both triplet and singlet superfields are present and in section \ref{lamWH} we present the result for the approximate convergence of the dimensionless parameters $\vec\lambda$ in the triplet/singlet extension of the MSSM. In section \ref{concl} we conclude.



\section{Minimal Supersymmetric Model}\label{secmssm}
\begin{figure}[t]
\centering
\mbox{\subfigure[]{
\includegraphics[width=0.48\textwidth]{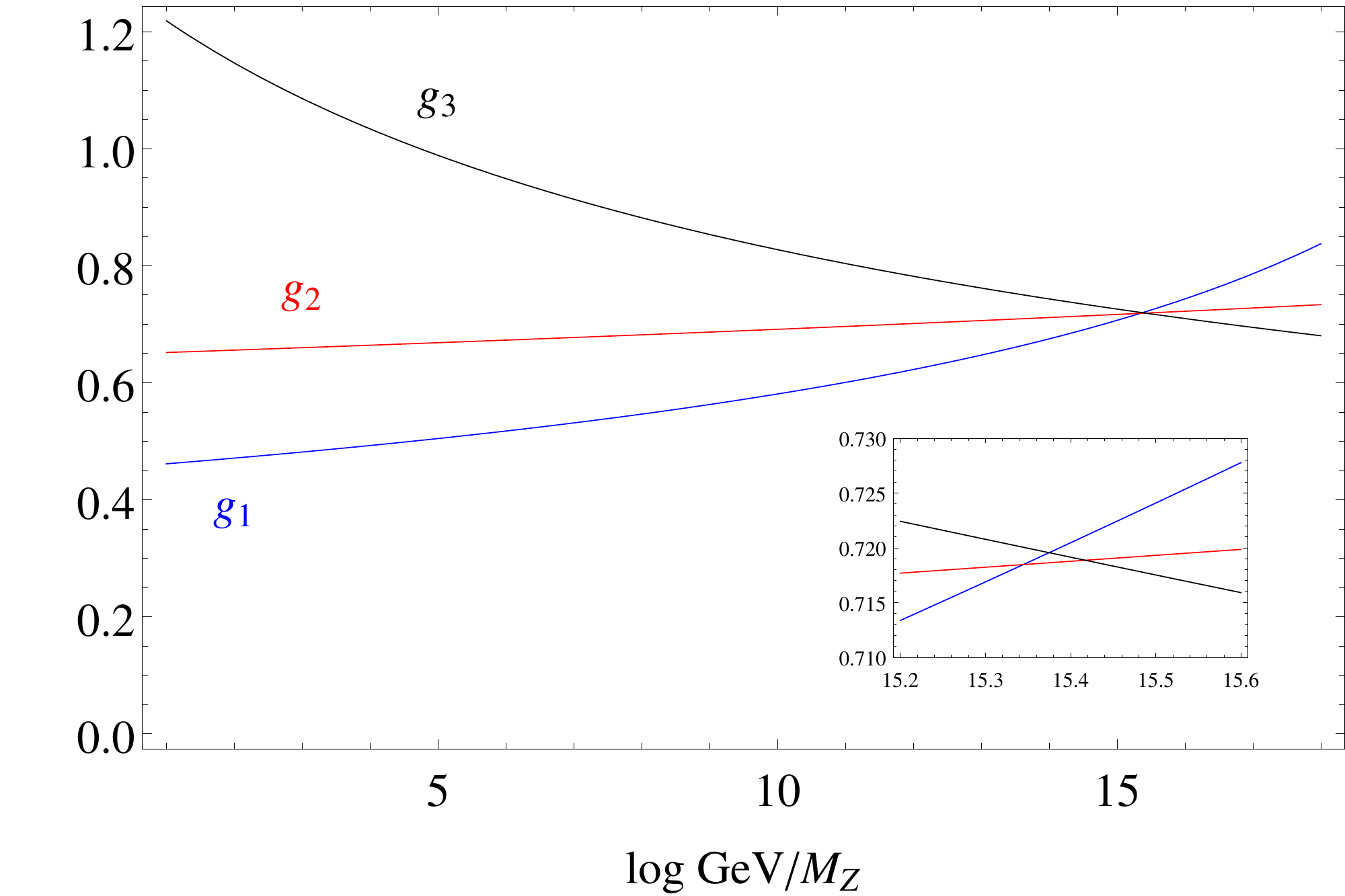}}
\subfigure[]{\includegraphics[width=0.48\textwidth]{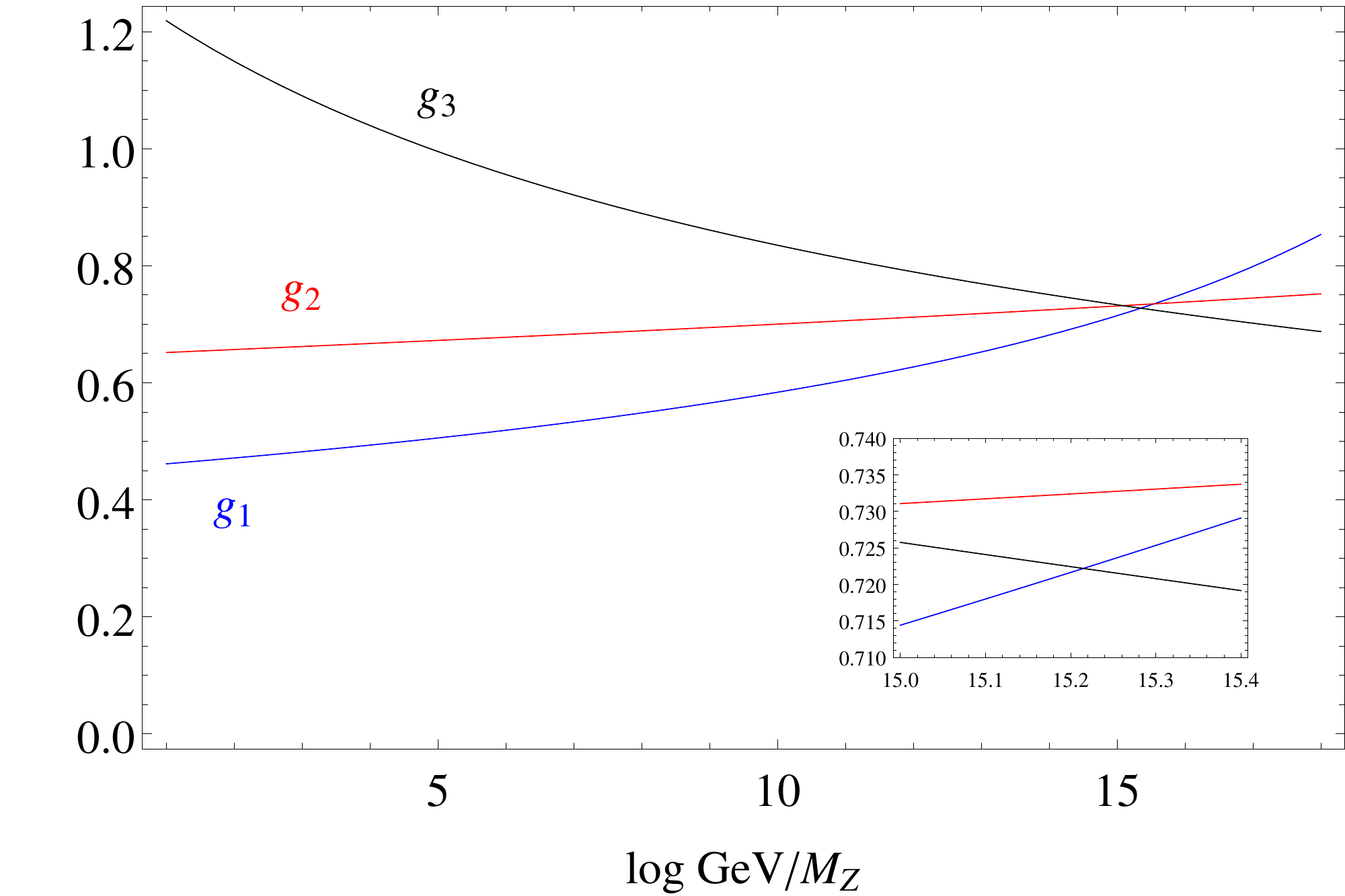}}}
\mbox{\subfigure[]{
\includegraphics[width=0.48\textwidth]{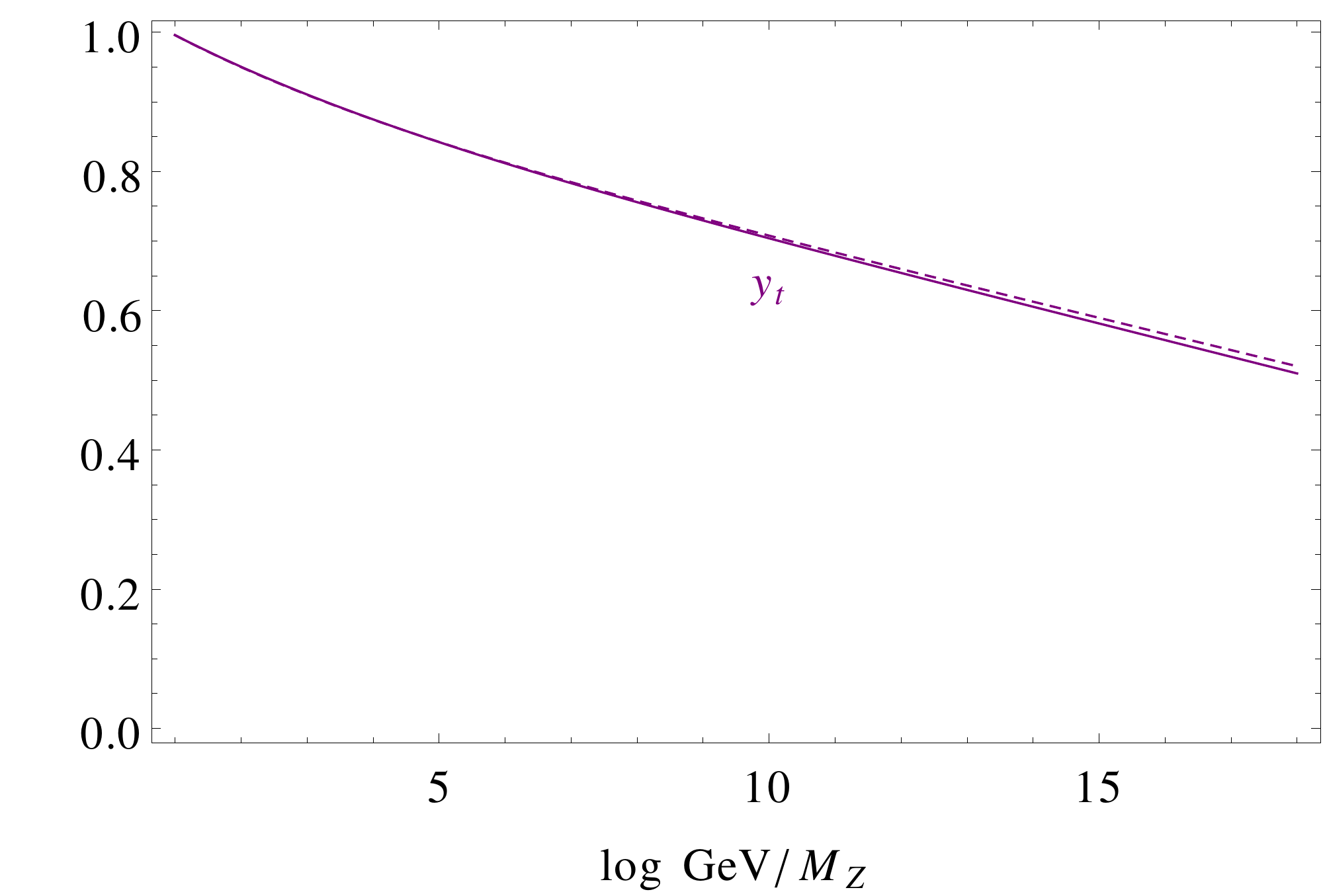}}}
\caption{Running of the gauge couplings in the MSSM with RG calculated at one-loop (a) and at two-loop (b). The one-loop (dashed line) and two-loop (solid line) running of the Yukawa couplings of the top quark (c).}
 \label{mssm}
\end{figure}

The Minimal Supersymmetric Standard Model (MSSM) is the supersymmetric version of the Standard Model (SM) in its minimal formulation. It contains a superfield for each field of the SM and an extra Higgs superfield, needed to construct an holomorphic superpotential. The superpotential of the MSSM is  given by
\bea
\hat{\mathcal{W}}_{MSSM}=y_u \hat U \hat H_u \hat Q - y_d \hat D \hat H_d \hat Q - y_e  \hat E \hat H_d \hat L + \mu \hat H_u \hat H_d
\eea
The RG equations at one-loop for the gauge couplings in the MSSM are
\bea\label{mssmrg}
\frac{d\, g_1}{dt}=\frac{33}{80\pi^2}\,g_1^3,\quad\frac{d\, g_2}{dt}=\frac{1}{16\pi^2}\,g_2^3,\quad\frac{d\, g_3}{dt}=-\frac{3}{16\pi^2}\,g_3^3
\eea
Their expressions at two-loop are reported in Appendix \ref{APPtl}, as well as the one- and two-loop expression for the dimensionless parameters considered in the following sections.
We present in Figure \ref{mssm} the behavior of the gauge couplings in the MSSM. We can see the well know result of the gauge couplings unification, shown in Figure \ref{mssm} (a), happening at $\sim10^{17}$ GeV. With the RG equations calculated at one-loop the gauge couplings unification is almost perfect, as we can appreciate in the zoomed region. In Figure \ref{mssm} (b) we show the, anyhow small, effect of the RG equations at two-loop. In the MSSM the only relevant coupling in the two-loop evolution of the gauge couplings is the Yukawa coupling of the top quark. We plot in Figure \ref{mssm} (c) its running at one-loop (dashed line) and at two-loop (solid line). 

The gauge couplings unification in the MSSM is a well known result and moreover the running of the dimensionless parameters is constrained by the electroweak value of the gauge and Yukawa couplings. It has been considered for the sake of completeness. The extensions of the MSSM discussed in the next sections will make the analysis far more interesting.

\section{The Simpliest Extension: NMSSM}\label{secnmssm}
There are many theoretical reasons to consider extensions of the MSSM \cite{MSSMextens} and the simplest one is the addition of a gauge singlet. The Next-to-Minimal Supersymmetric Standard Model (NMSSM) is well known for the elegant solution to the $\mu$ problem of the MSSM, as well as for many other theoretical and phenomenological implications \cite{NMSSM}. The superpotential of the model is\footnote{We consider the $ Z_3$ symmetric version of NMSSM because we are interested in the running of the dimensionless couplings.}
\bea\label{WNMSSM}
\hat{\mathcal{W}}_{NMSSM}=y_u \hat U \hat H_u \hat Q - y_d \hat D \hat H_d \hat Q - y_e  \hat E \hat H_d \hat L + \lambda \hat S \hat H_u \hat H_d +\kappa \hat S^3
\eea
Because of the addition of a superfield uncharged under the gauge group $SU(3)_c\times SU(2)_w\times U(1)_Y$, the one-loop expressions of the RG equations for the gauge couplings in the NMSSM are exactly the same as in the MSSM, \textit{cf.} Eq. \ref{mssmrg}. 
However at two-loop the situation is quite different. In fact in the MSSM the only dimensionless couplings present are the gauge couplings and the Yukawa couplings. In the NMSSM there are two more dimensionless parameters, namely $\lambda$ and $\kappa$, as we can see from Eq. \ref{WNMSSM}. These two dimensionless couplings enter in the two-loop expression of the RG equations giving rise to a system of coupled differential equations.
\begin{figure}[t]
\centering
\mbox{\subfigure[]{
\includegraphics[width=0.48\textwidth]{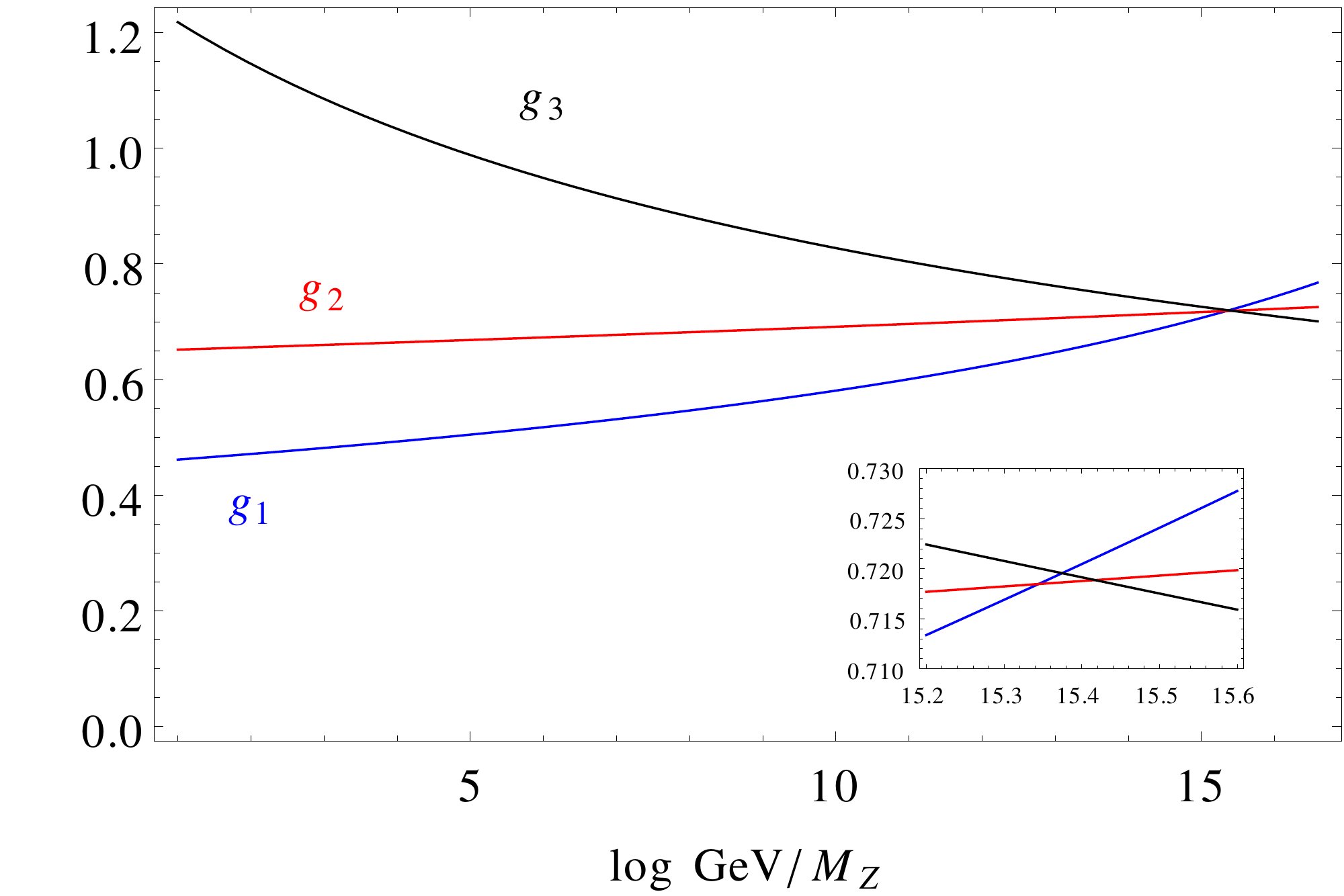}}
\subfigure[]{\includegraphics[width=0.48\textwidth]{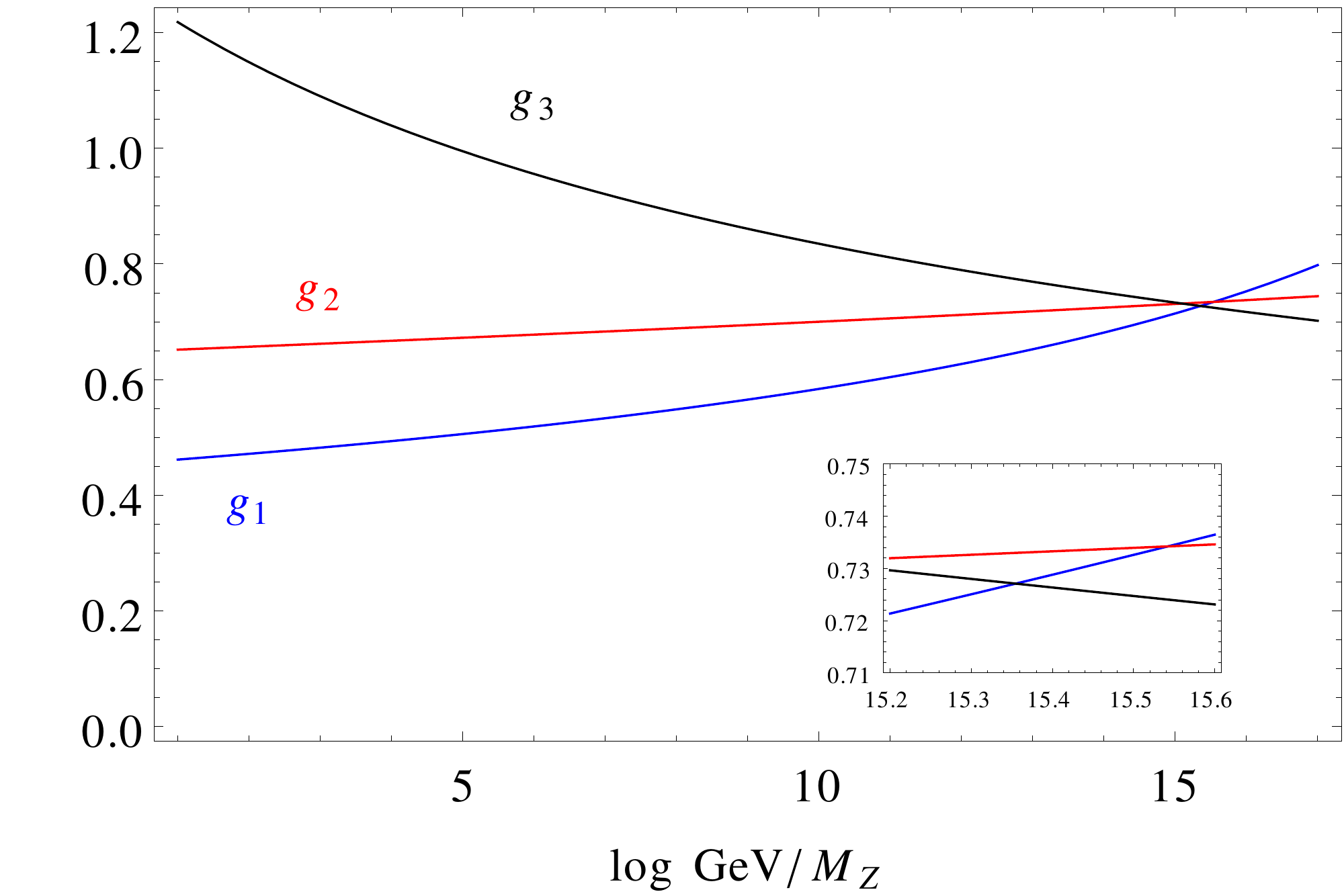}}}
\mbox{\subfigure[]{
\includegraphics[width=0.48\textwidth]{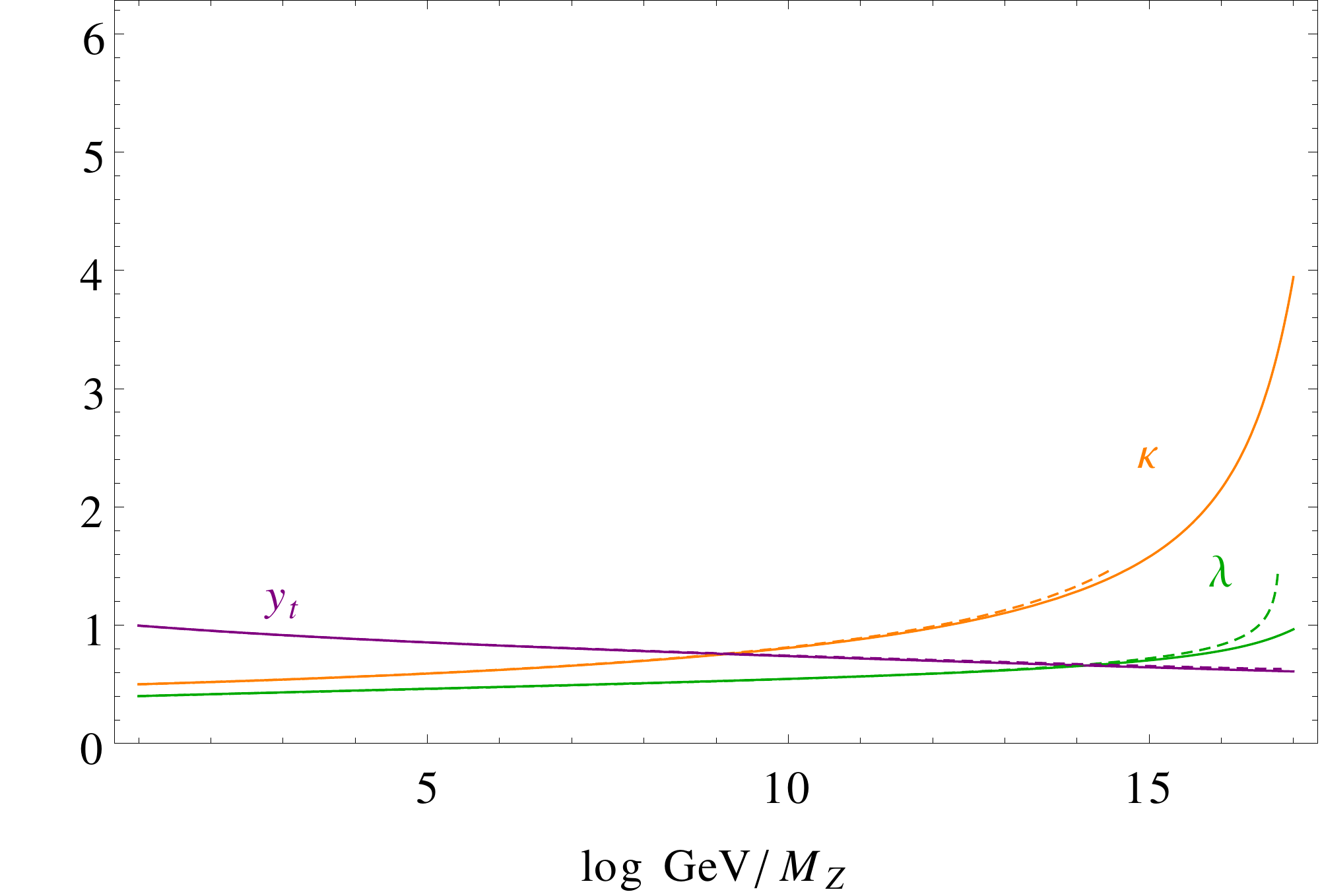}}}
\caption{Running of the gauge couplings in the NMSSM with RG calculated at one-loop (a) and at two-loop (b). The one-loop (dashed line) and two-loop (solid line) running of the Yukawa couplings of the top quark and for $\kappa$ and $\lambda$ (c).}
 \label{nmssmlow}
\end{figure}

In Figure \ref{nmssmlow} we plot the running of the gauge couplings at one-loop (Figure \ref{nmssmlow} (a)) and at two-loop (Figure \ref{nmssmlow} (b)) as well as the running of $\lambda$, $\kappa$ and the Yukawa coupling of the top quark (Figure \ref{nmssmlow} (c)). In Figure \ref{nmssmlow} (c) the dashed lines are the one-loop curves whereas the solid lines represent the two-loop ones. A close inspection of Figure \ref{mssm} (a) and Figure \ref{nmssmlow} (a) reveal that at one-loop the situation is exactly the same for MSSM and NMSSM. This is obvious because, as already stated, the singlet superfield is uncharged under the gauge group and hence the equations governing the evolution of the gauge couplings at one-loop are exactly the same.

At two-loop the running of the gauge couplings slightly changes, as we can see in Figure \ref{nmssmlow} (b). However the most interesting difference between the one- and the two-loop is the behavior of the other dimensionless couplings, in particular $\kappa$ and $\lambda$. In fact in Figure \ref{nmssmlow} (c) we can see the appearance of a Landau pole at $\sim10^{18}$ GeV at one-loop (dashed line). The singularity is not present at two-loop and the couplings remain perturbative until the Planck scale, as showed by the solid lines of Figure \ref{nmssmlow} (c).
\begin{figure}[t]
\centering
\mbox{\subfigure[]{
\includegraphics[width=0.49\textwidth]{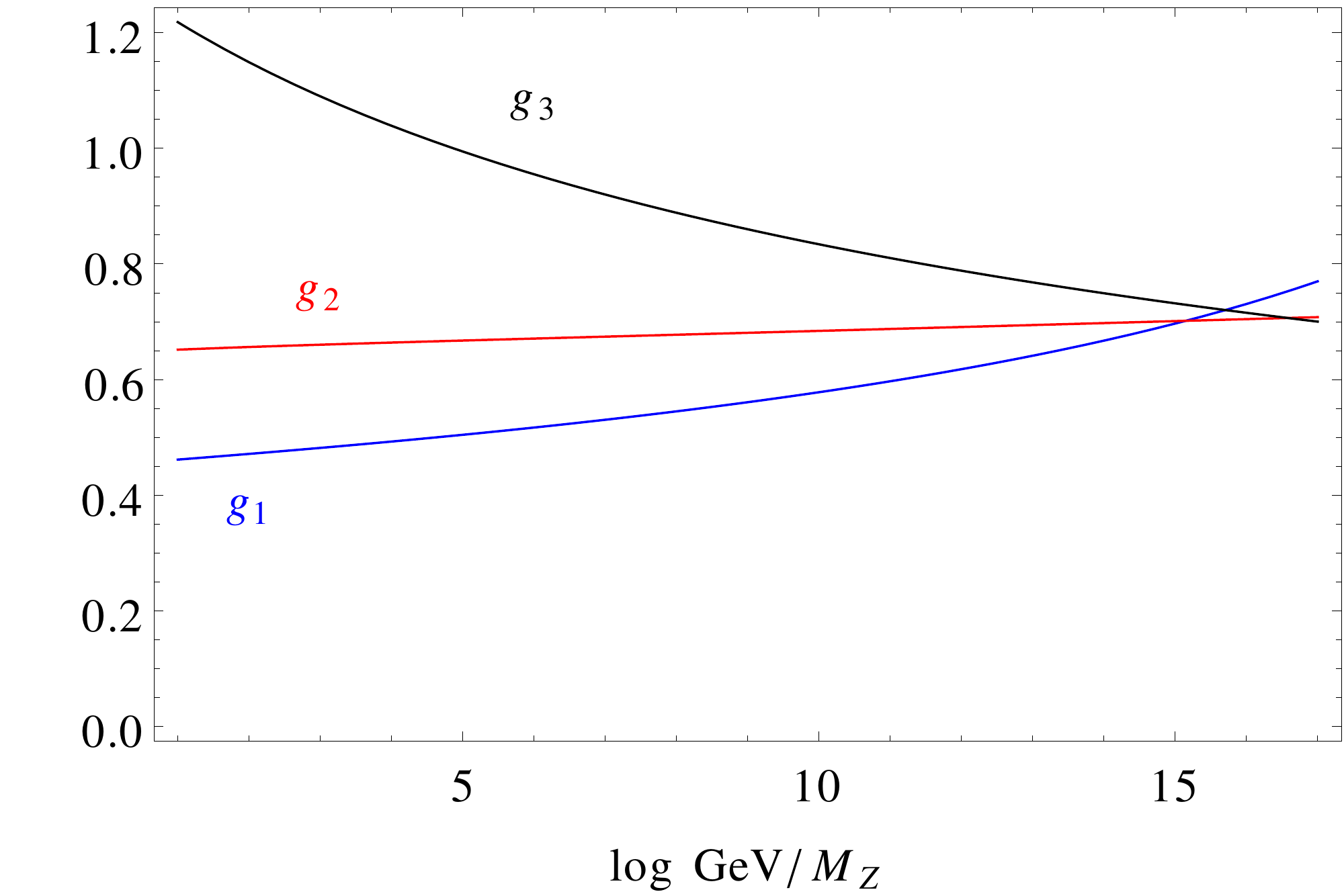}}
\subfigure[]{\includegraphics[width=0.48\textwidth]{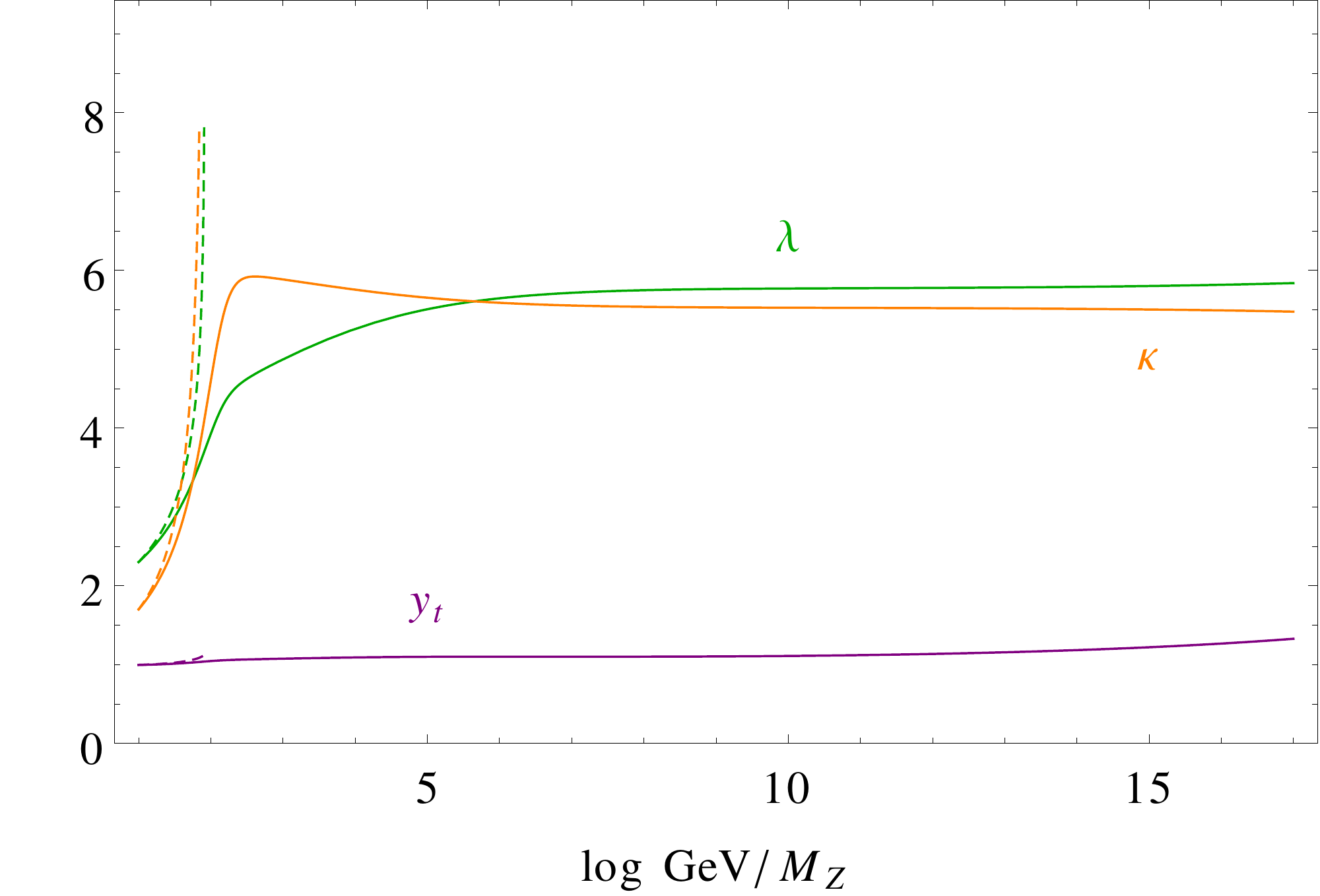}}}
\caption{Running of the gauge couplings in the NMSSM with RG calculated at two-loop (a) and the one-loop (dashed line) and two-loop (solid line) running of the Yukawa couplings of the top quark and for $\kappa$ and $\lambda$ (b).}
 \label{nmssmmid}
\end{figure}

In Figure \ref{nmssmmid} we present a different situation. In Figure \ref{nmssmmid} (a) we plot the running of the gauge couplings at two-loop and in Figure \ref{nmssmmid} (b) we plot the one and two-loop running of the relevant dimensonless couplings. At one-loop the dimensionless couplings became divergent at relatively low energies ($\sim10^4$ GeV). This happens because their starting values are quite large ($\lambda\sim\kappa\sim2$). However their two-loop evolution present no singularity and the couplings are well within the perturbative range ($4\pi$) until the Planck scale. 

The two-loop RG equations do not present the appearance of a Landau pole for a wide range of values of the dimensionless couplings at the electroweak scale. This is particularly evident when we deal with $\lambda,\,\kappa \gsim 1$. In these situations the running of the dimensionless couplings, calculated with RG at one-loop, present the appearance of a Landau pole at relatively low energies, $\lesssim10^5$ GeV. This behavior disappear using the RG equation at two-loop. This is due to the similar value of the one- and two-loop contribution to the RG equations when the electroweak values of the dimensionless parameters is large enough. We will exploit this result in the next section in connection with the gauge couplings unification.

\section{Addition of Triplet Superfields}\label{sectriplet}
The addition to the MSSM of triplet superfield(s) generally spoils the gauge couplings unification. This is a well known results and it is due to the fact that at one-loop the RG equation for each gauge coupling of $SU(3)_c\times SU(2)_w\times U(1)_Y$ is indipendent from any other parameter of the theory
\bea\label{gengauge}
\frac{d g_i(t)}{dt}=c_{g_i} g^3_i(t)
\eea
and the $c_{g_i}$ are related to the field content of the model. If one introduces triplet(s) superfield(s) then $c_{g_1}$ and $c_{g_2}$ are such that the MSSM-like unification of the gauge couplings is lost. The possibility of a different kind of gauge couplings unification in models with triplet(s) has been considered in \cite{tripnouni}.

Eq. \ref{gengauge} is a one-loop result whereas, as we have already pointed out, at two-loop the RG equations for the dimensionless parameters form a set of coupled differential equations. We have seen in section \ref{secnmssm} that there is no drastic change in the running of the gauge couplings at one- or two-loop in the case of NMSSM, the reason being the singlet, thus uncharged, superfield added to the MSSM superfield content. In theories with triplet(s) the situation can be very different. The two-loop RG equations for the gauge couplings, in particular the $g_2$ one, depends directly on all the parameter of the triplet's interactions. The competition between one- and two-loop contributions to the RG equations can be more relevant if $\vec\lambda\gsim1$, leading to unexpected results about the gauge couplings unification. 

\subsection{TMSSM}
We consider the simplest extension of the MSSM including a superfield in the triplet representation of $SU(2)_w$ \cite{tmssmpaper}, that is a $Y=0$ triplet superfield in addition to the superfield content of the MSSM. The superpotential of the model is
\bea\label{WTMSSM}
\hat{\mathcal{W}}_{TMSSM}=y_u \hat U \hat H_u \hat Q - y_d \hat D \hat H_d \hat Q - y_e  \hat E \hat H_d \hat L + \mu \hat H_u \hat H_d + M_T \hat T^2 + \lambda \hat H_u\hat T \hat H_d 
\eea
Obviously the RG equations for the gauge couplings are different from the MSSM case. Their one-loop expression is
\bea\label{tmssmrg}
\frac{d\, g_1}{dt}=\frac{33}{80\pi^2}\,g_1^3,\quad\frac{d\, g_2}{dt}=\frac{3}{16\pi^2}\,g_2^3,\quad\frac{d\, g_3}{dt}=-\frac{3}{16\pi^2}\,g_3^3
\eea
The only difference is in the RG equation for $g_2$ because $\hat T\in(\bf1,\bf3,\bf0)$ of $SU(3)_c\times SU(2)_w\times U(1)_Y$.

\begin{figure}[t]
\centering
\mbox{\subfigure[]{
\includegraphics[width=0.49\textwidth]{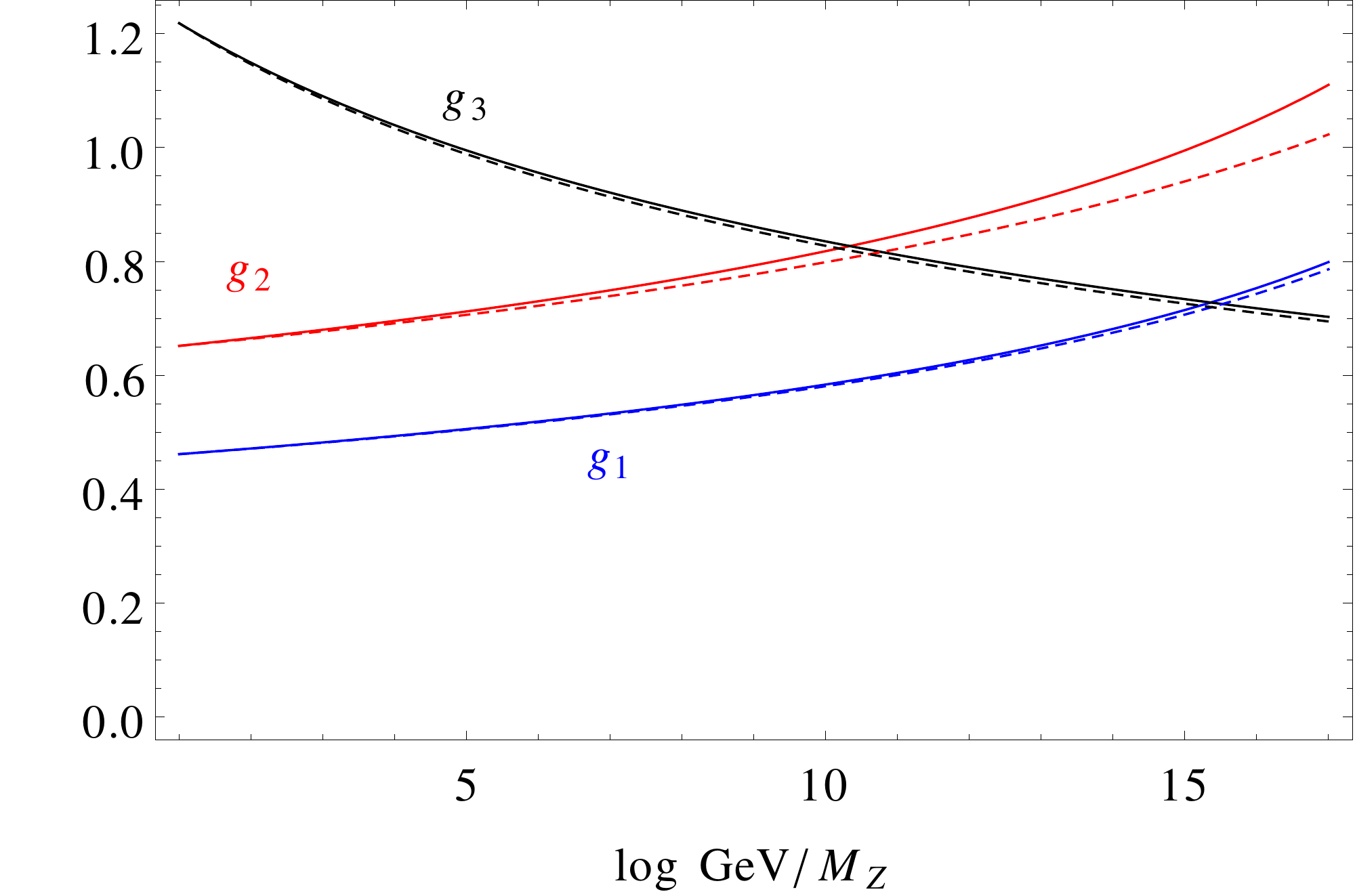}}
\subfigure[]{\includegraphics[width=0.48\textwidth]{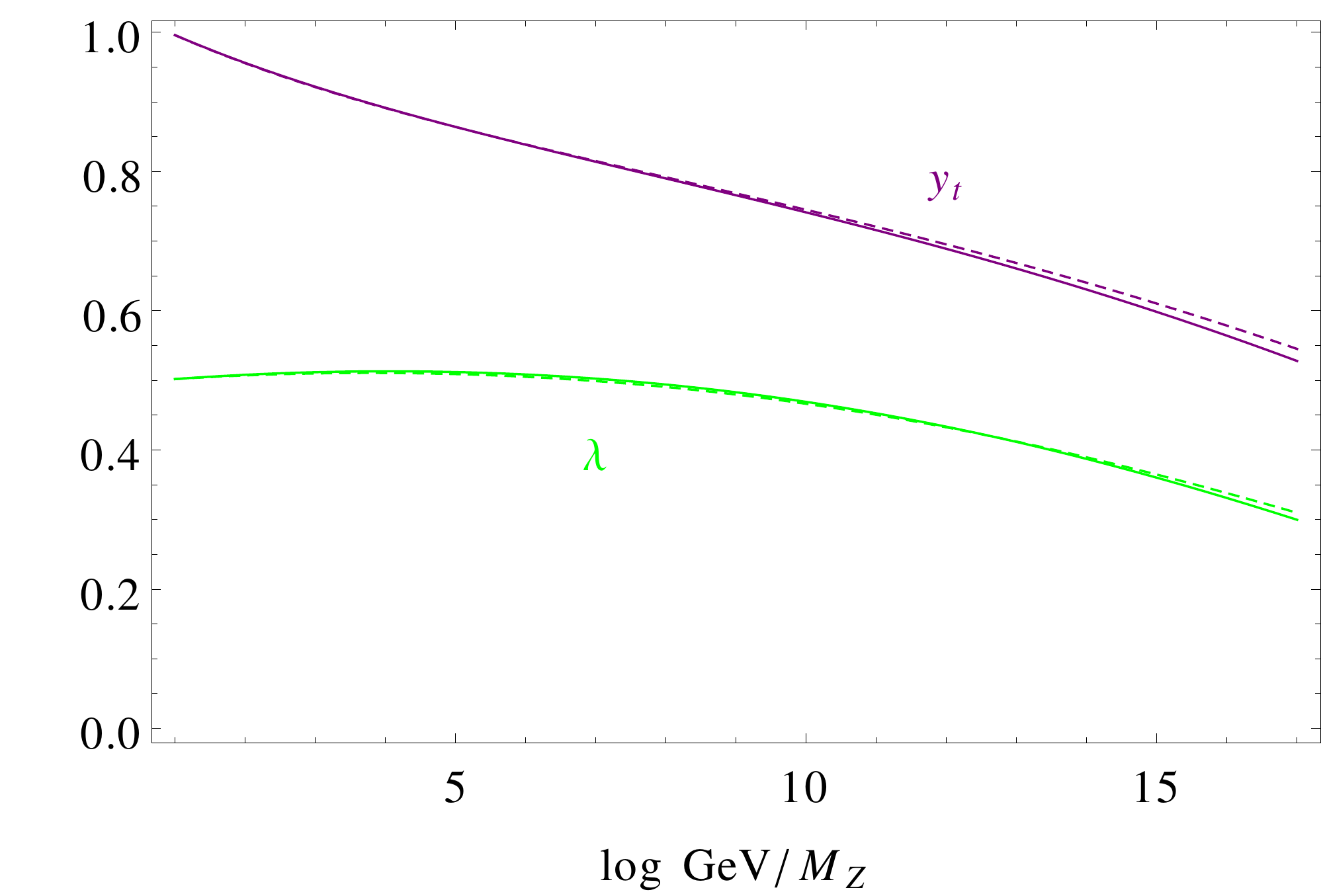}}}
\mbox{\subfigure[]{
\includegraphics[width=0.49\textwidth]{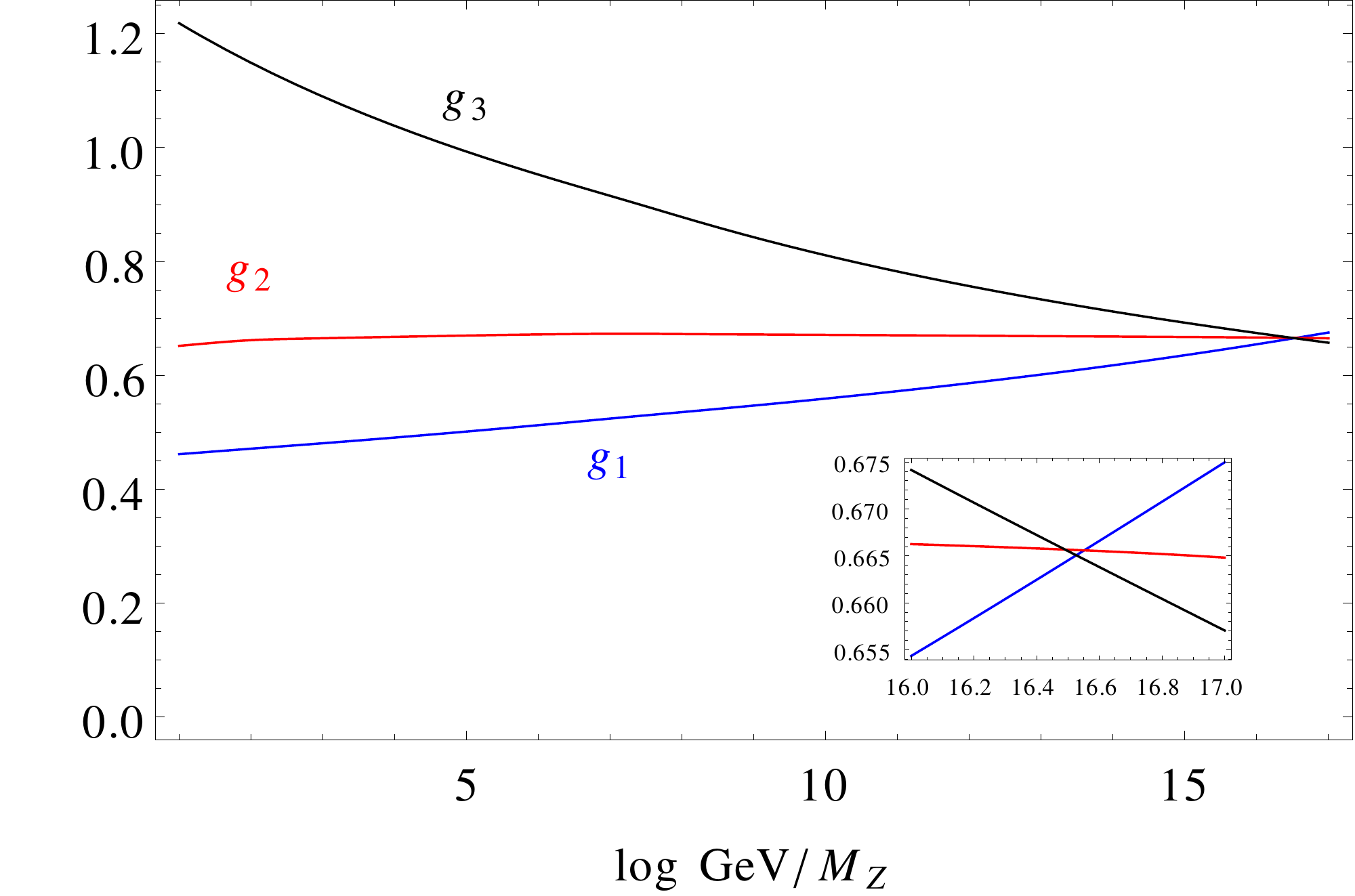}}
\subfigure[]{\includegraphics[width=0.48\textwidth]{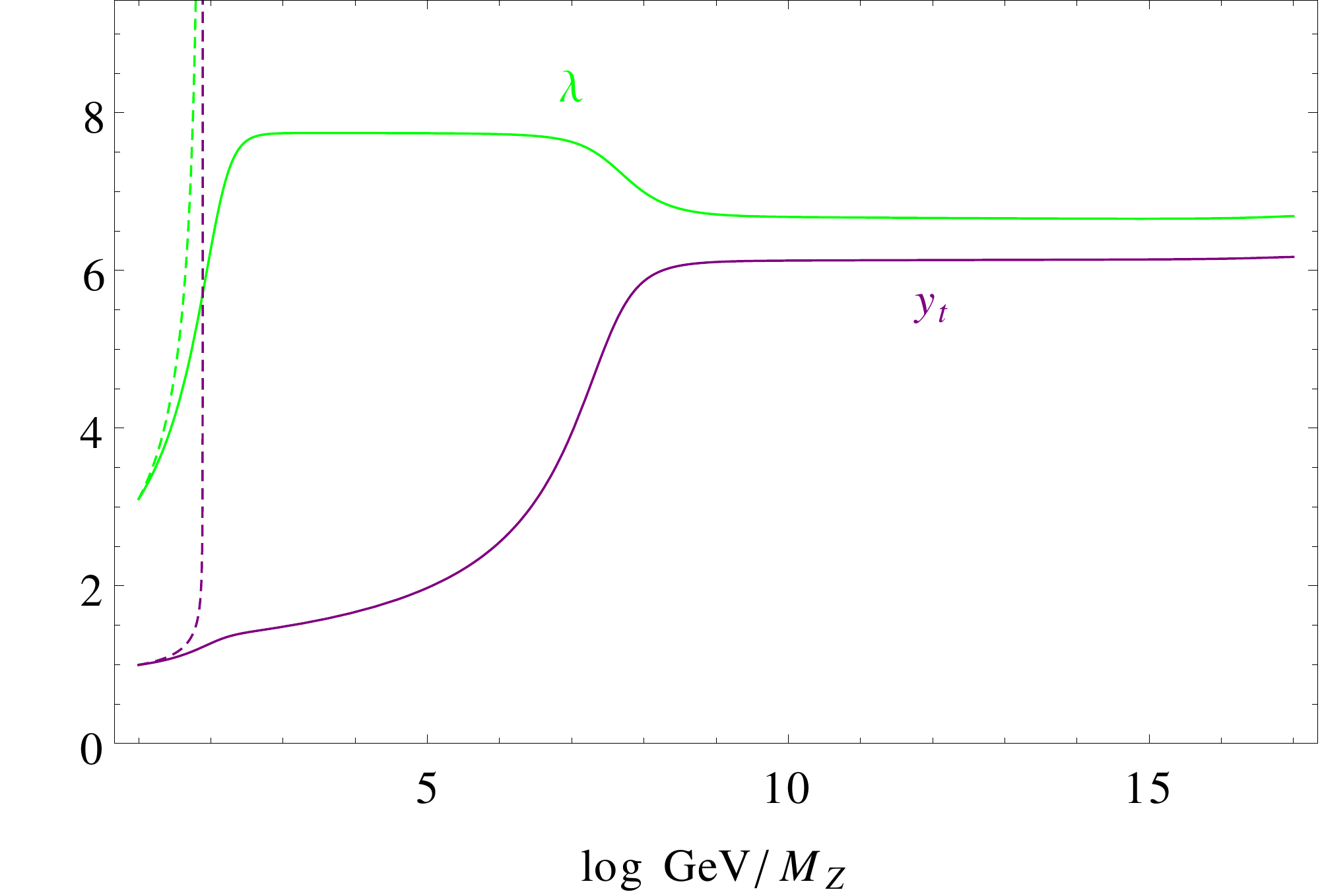}}}
\caption{Running of the gauge couplings in the TMSSM with RG calculated at two-loop (a)-(c) and the one-loop (dashed line) and two-loop (solid line) running of the Yukawa couplings of the top quark and $\lambda$ (b)-(d).}
 \label{tmssm}
\end{figure}

In Figure \ref{tmssm} (a)-(b) we plot the running of the gauge couplings in the TMSSM model as well as the running of the Yukawa coupling of the top quark and the dimensionless coupling $\lambda$, both at one- and at two-loop order. The dashed and solid lines in Figure \ref{tmssm} (a)-(b) represent the evolution of the couplings at one- and two-loop order respectively. The parameters remain in the perturbative range until the Planck scale but the gauge couplings do not unify neither at one-loop nor at two-loop. This happens because of the different coefficient of the $\beta$-function of $g_2$ in Eq. \ref{mssmrg} and Eq. \ref{tmssmrg}. We can see that the difference in the one- and two-loop evolution of the dimensionless couplings is quite small. This situation is similar to the MSSM and NMSSM case, apart from the unification of $g_1$, $g_2$ and $g_3$.

We now consider a scenario where $\lambda>1$ at the electroweak scale. As we discussed in the previous section, if the electroweak value of a dimensionless coupling is greater than one usually the coupling will diverge for some $M < M_{\rm{Planck}}$ when the RG equation governing its evolution is calculated at one-loop. However the condition $\lambda>1$ at the electroweak scale is sufficient to change the two-loop evolution of the dimensionless parameters, as we have already shown. Instead of the appearance of a Landau pole, $\lambda$ and $y_t$ will approach a UV fixed point in their evolution. This have an unexpected impact on the running of the gauge couplings and can possibly restore the gauge couplings unification below the Planck scale.

This is show in Figure \ref{tmssm} (c)-(d) where we plot the running of the gauge couplings with the RG equations calculated at two-loop, Figure \ref{tmssm} (c), and the running of $\lambda$ and $y_t$ at one-loop (dashed lines) and at two-loop (solid lines), Figure \ref{tmssm} (d). Similarly to the NMSSM case, the two-loop running of the couplings present no Landau pole until the Planck scale. Conversely at $\sim10^{10}$ GeV $\lambda$ and $y_t$ approach a UV fixed point. The couplings are, in any case, within the perturbative range ($4\pi$) until the Planck scale. In particular in Figure \ref{tmssm} (c)-(d) the electroweak value of $\lambda$ is such that the gauge couplings unification is achieved at two-loop, as shown in the zoomed plot of Figure \ref{tmssm} (c).  

\subsection{TcMSSM}
Another extension of the MSSM including superfields in the triplet representation is the Triplet-custodial MSSM (TcMSSM). It has three triplet superfields on the top of the MSSM superfield content. One of the triplet, $\hat T$, has vanishing hypercharge whereas $\hat t$ and $\hat{\bar t}$ have $Y=1$ and $Y=-1$ respectively. We have dubbed it TcMSSM because in this model the custodial symmetry at tree level can be naturally imposed requesting the alignment of the vacuum expectation values of the triplets \cite{quiros,quiros2}. The superpotential of the model is
\begin{align}\label{Wtcmssm}
\hat{\mathcal{W}}_{\rm{TcMSSM}}&=y_u \hat U \hat H_u \hat Q - y_d \hat D \hat H_d \hat Q - y_e  \hat E \hat H_d \hat L + \mu \hat H_u \hat H_d + M_T \hat T^2 + M_t \hat t \hat{\bar t}\nn\\
&\quad + \xi_u \hat H_u\,\hat{\bar t}\,\hat H_u +\xi_d \hat H_d\,\hat t\,\hat H_d + \lambda \hat H_u\,\hat T\,\hat H_d + \zeta\, \hat t\, \hat T\, \hat{\bar t}
\end{align}
The model has a global $SU(2)_L\times SU(2)_R$ symmetry which is spontaneously broken to the $SU(2)_V$ custodial symmetry after electroweak symmetry breaking.

The one-loop expression for the RG equations of the gauge couplings are
\bea\label{tcmssmrg}
\frac{d\, g_1}{dt}=\frac{51}{80\pi^2}\,g_1^3,\quad\frac{d\, g_2}{dt}=\frac{7}{16\pi^2}\,g_2^3,\quad\frac{d\, g_3}{dt}=-\frac{3}{16\pi^2}\,g_3^3
\eea

\begin{figure}[ht]
\centering
\mbox{\subfigure[]{
\includegraphics[width=0.49\textwidth]{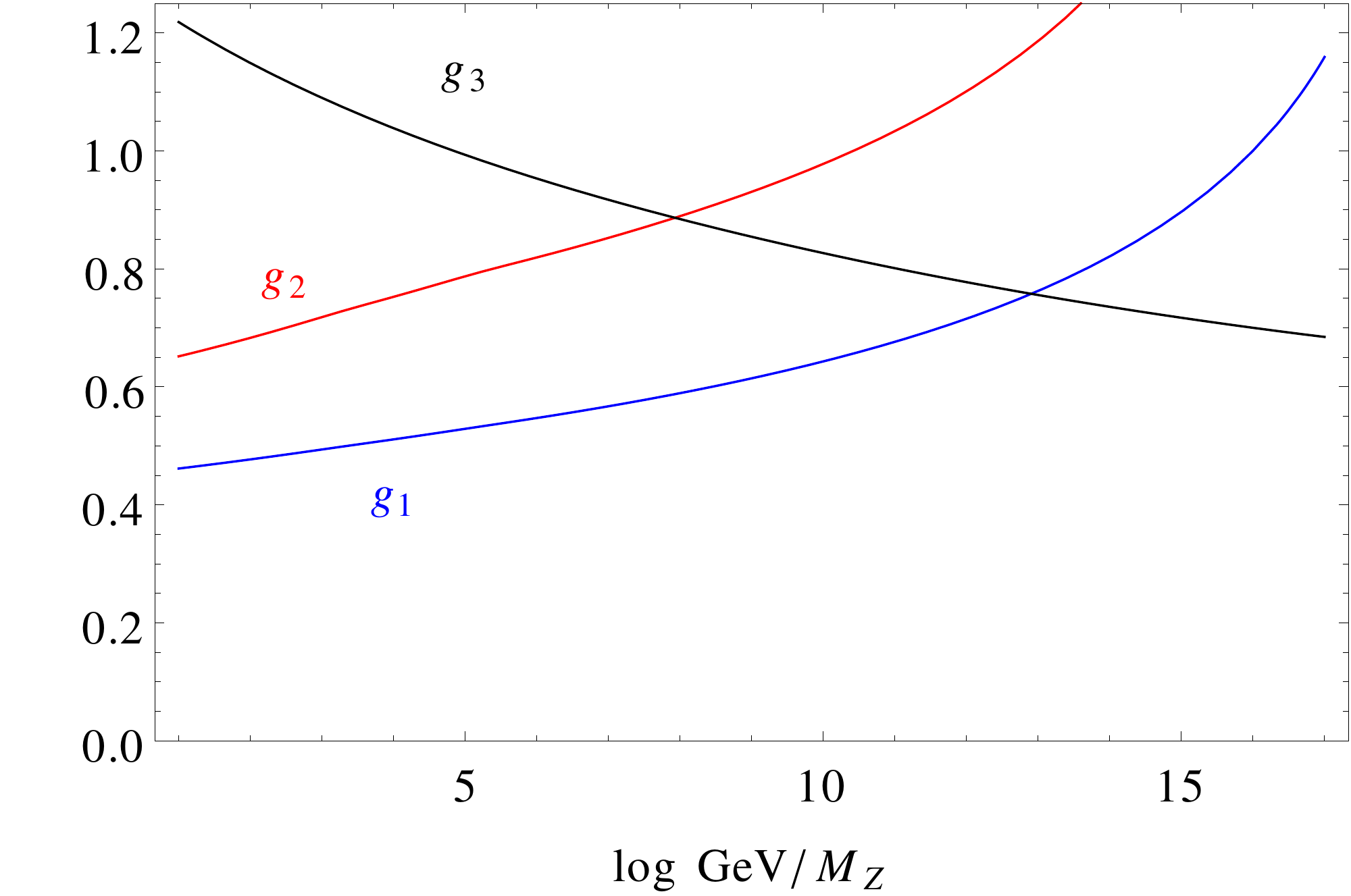}}
\subfigure[]{\includegraphics[width=0.48\textwidth]{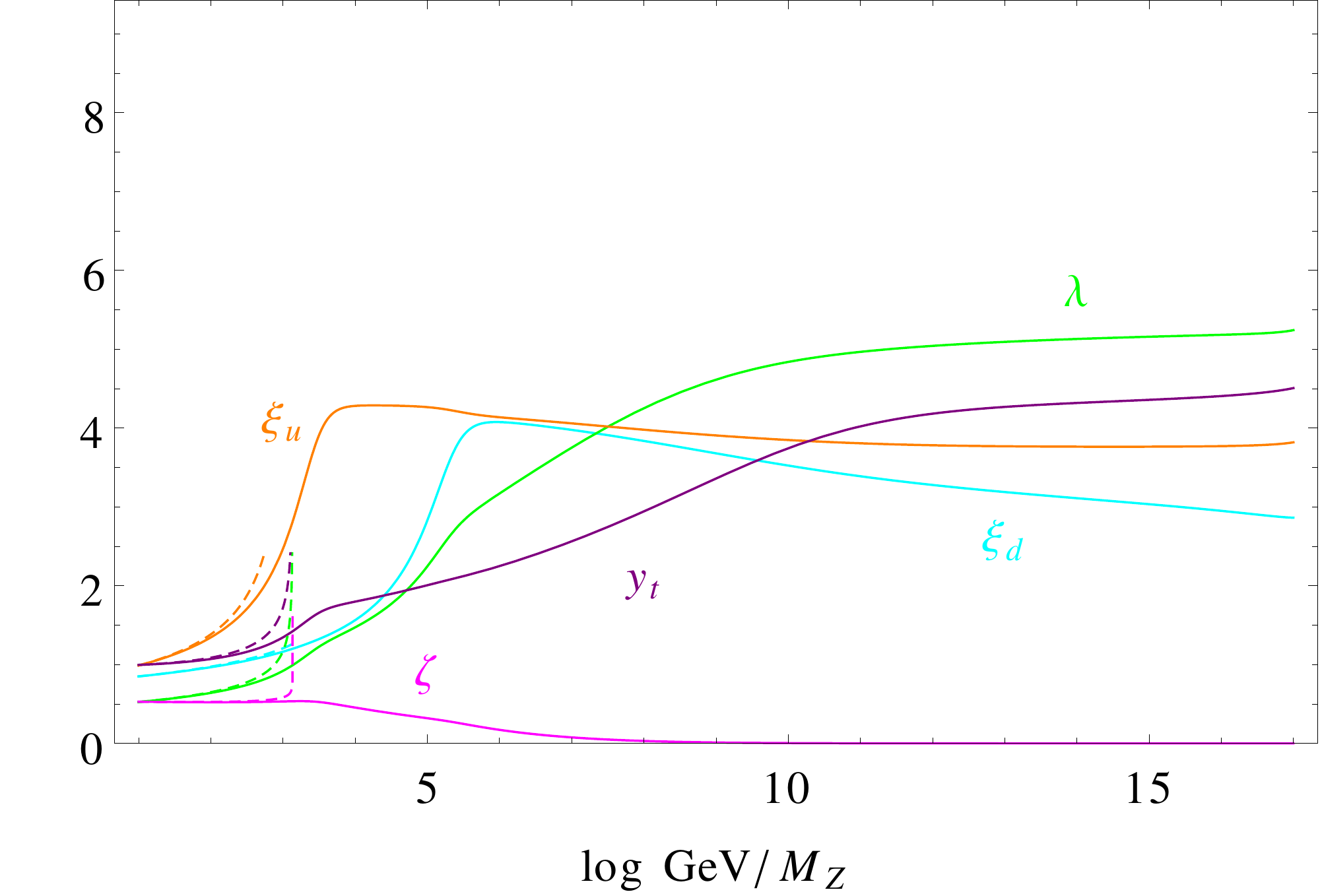}}}
\mbox{\subfigure[]{
\includegraphics[width=0.49\textwidth]{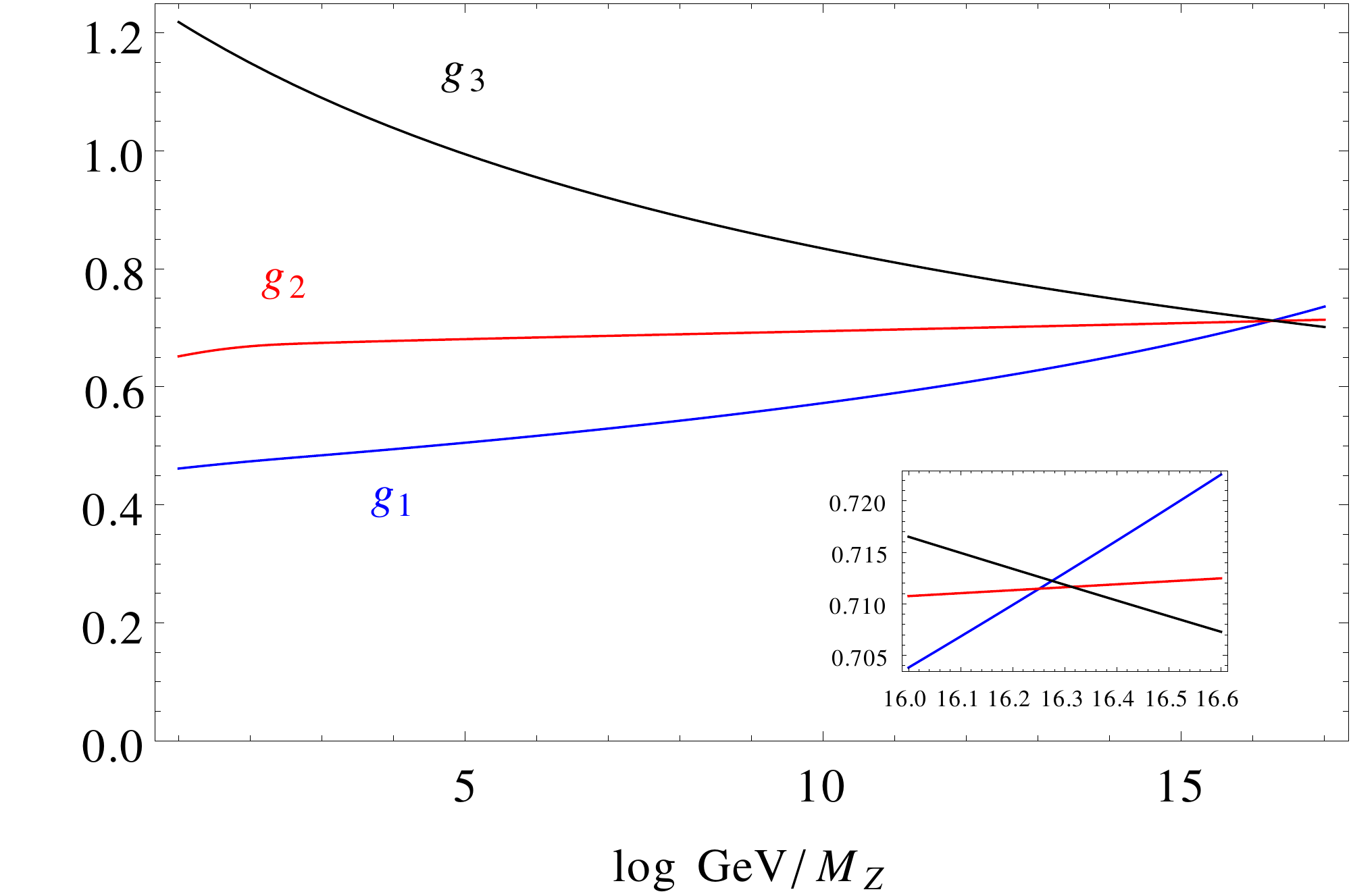}}
\subfigure[]{\includegraphics[width=0.48\textwidth]{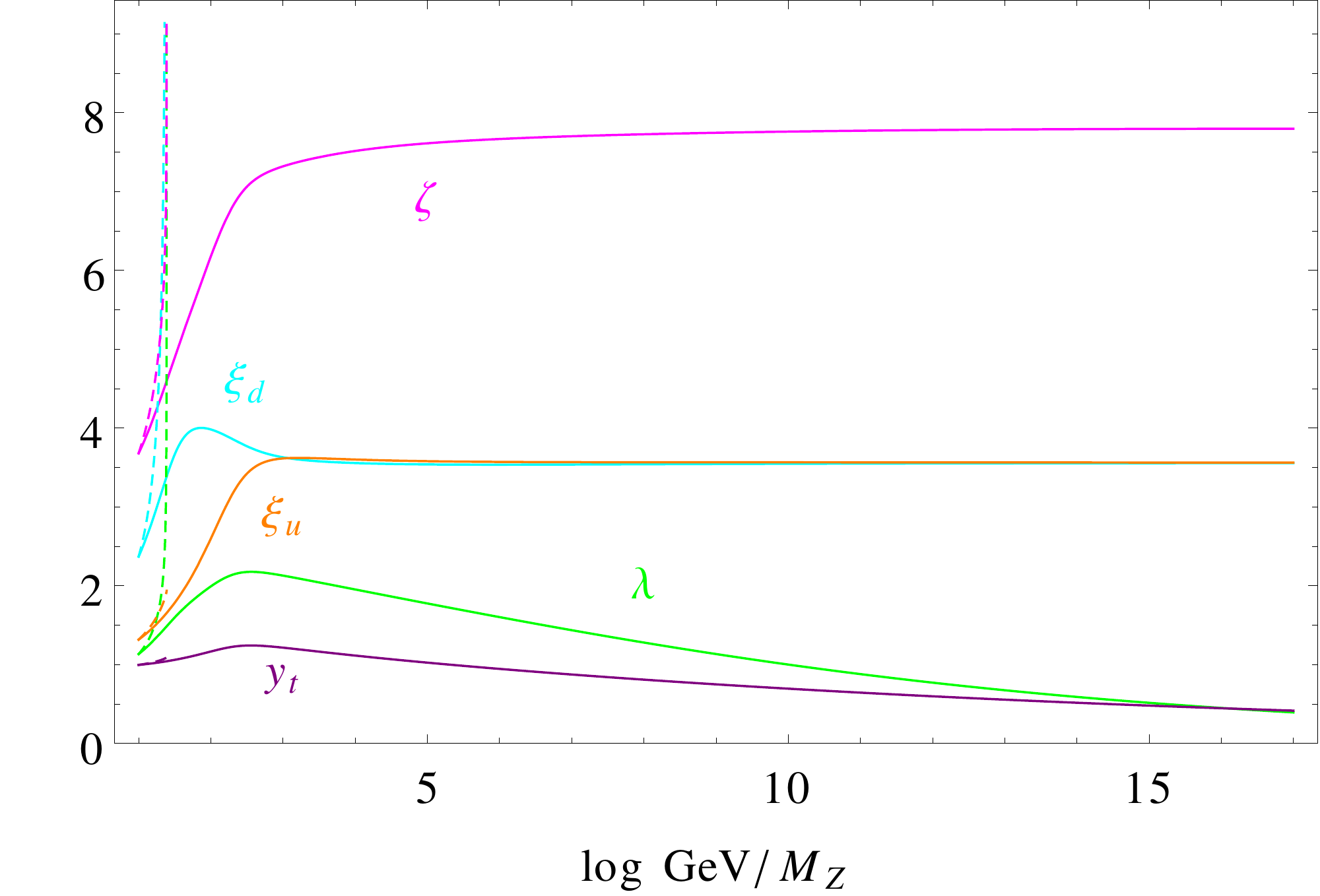}}}
\caption{Running of the gauge couplings in the TcMSSM with RG calculated at two-loop (a)-(c) and the one-loop (dashed line) and two-loop (solid line) running of the dimensionless couplings (b)-(d).}
 \label{tcmssmuni}
\end{figure}

In Figure \ref{tcmssmuni} (a)-(b) we plot the running of the dimensionless couplings in the TcMSSM. Similarly to the previous case, we can see that there is no gauge couplings unification (at two-loop) when $\lambda\lesssim1$, $\zeta\lesssim1$, $\xi_u\lesssim1$ and $\xi_d\lesssim1$. 

As shonw in Figure \ref{tcmssmuni} (c)-(d) the gauge couplings unification can be achieved at two-loop in the TcMSSM if the electroweak values of the dimensionless couplings is large enough. Conversly to the TMSSM case, where the two-loop unification is tuned by $\lambda$ and $y_t$, here we have more dimensionless couplings, so the situation is more flexible. In Figure \ref{tcmssmuni} (d) the dimensionless couplings approach a UV fixed point around the unification scale. Again we stress the fact that all the dimensionless couplings are within the perturbative range until the Planck scale.

\section{Triplet and Singlet Extension of the MSSM}\label{sectnssm}

\begin{figure}[t]
\centering
\mbox{\subfigure[]{
\includegraphics[width=0.49\textwidth]{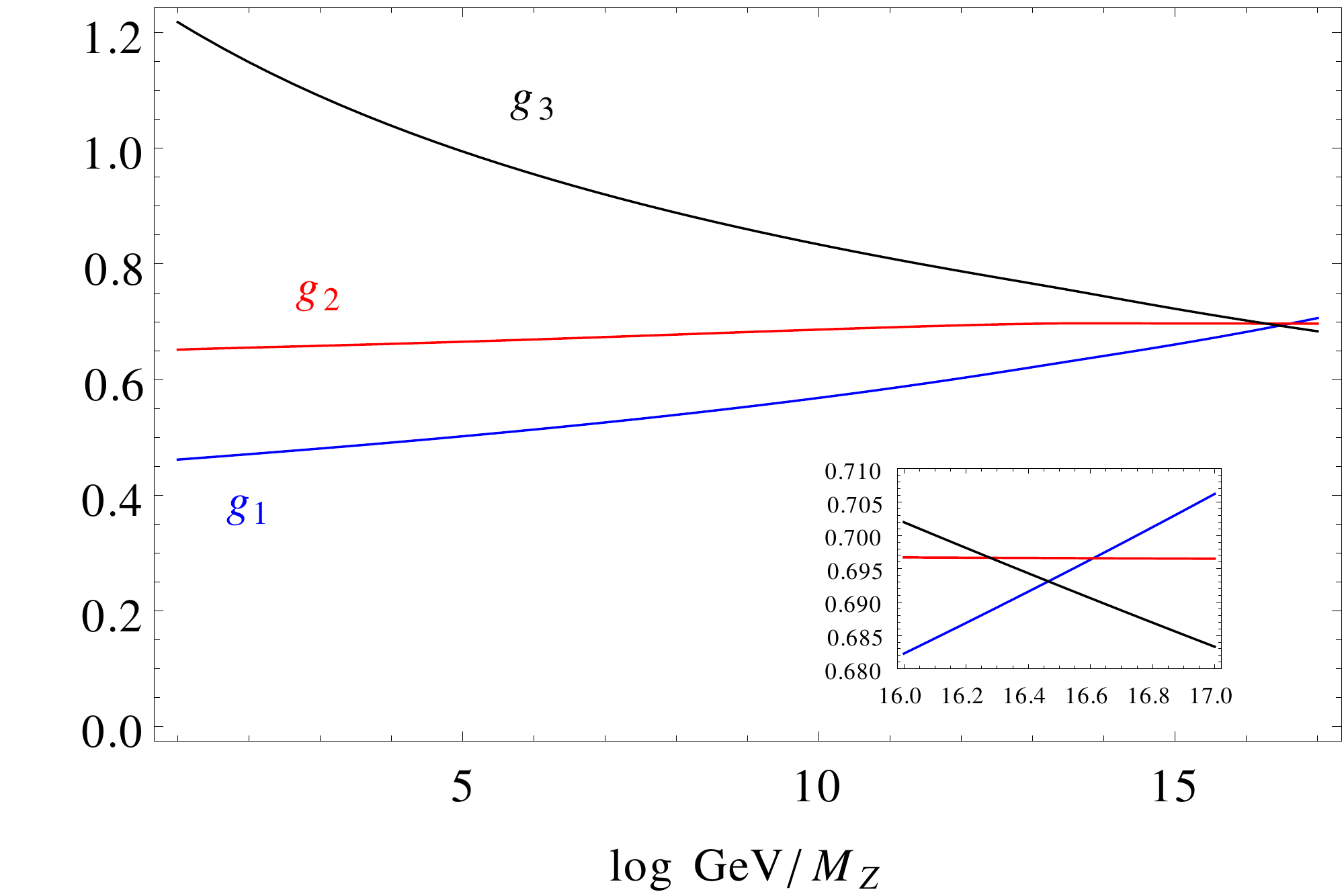}}
\subfigure[]{\includegraphics[width=0.48\textwidth]{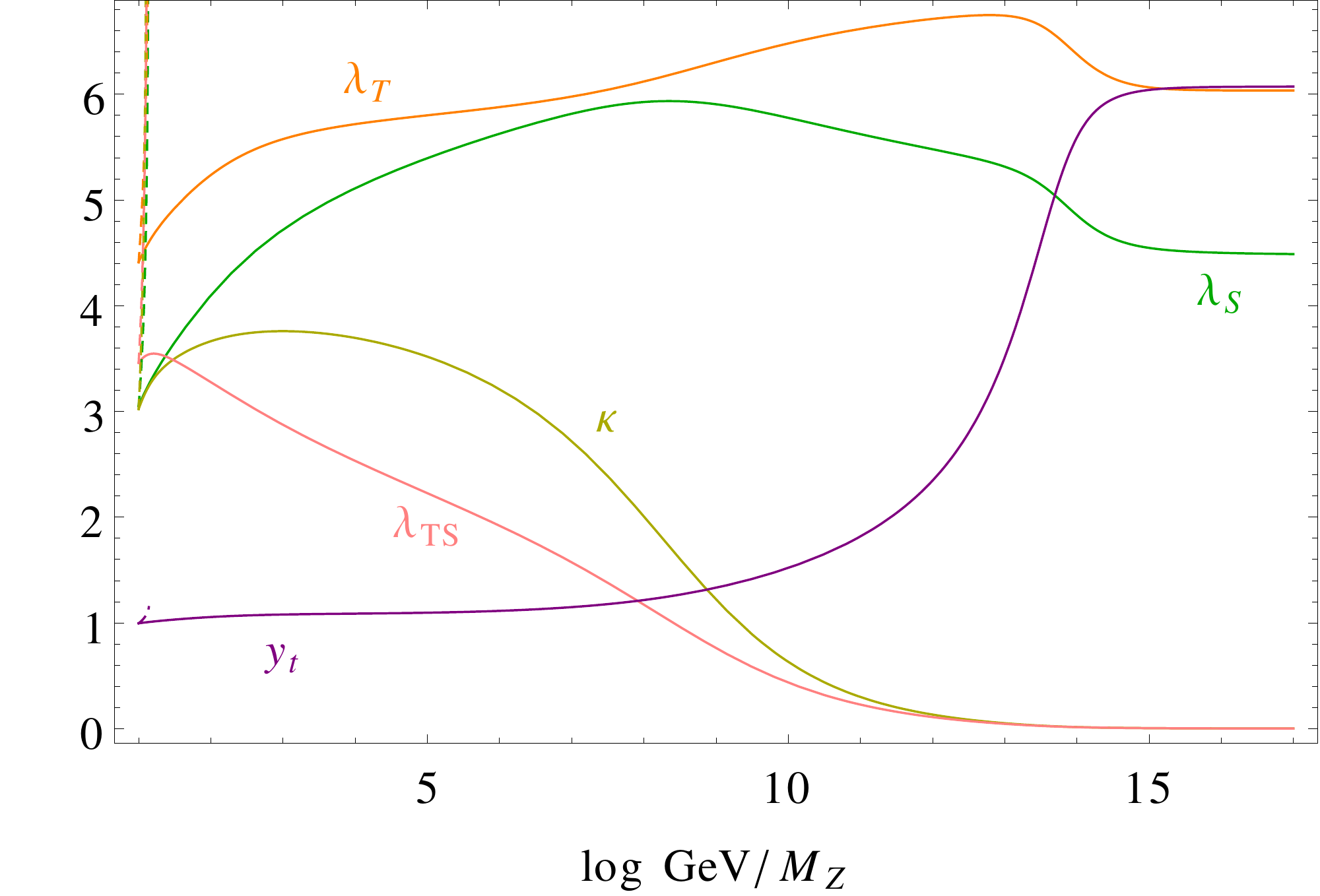}}}
\caption{Running of the gauge couplings in the TNMSSM with RG calculated at two-loop (a) and the one-loop (dashed line) and two-loop (solid line) running of the dimensionless couplings (b).}
 \label{tnssmuniH}
\end{figure}

\begin{table}
\begin{center}
\renewcommand{\arraystretch}{1.4}
\begin{tabular}{|c|c|c|c|}
\hline\hline
$m_{h_1}\sim125$ GeV&$m_{h_2}=8.920$ TeV&$m_{h_3}=9.475$ TeV&$m_{h_4}=31.96$ TeV\\
\hline
$m_{a_1}=21.35$ GeV & $m_{a_2}=9.266$ TeV& $m_{a_3}=32.87$ TeV&\multicolumn{1}{c}{}\\
\cline{1-3}
$m_{h^\pm_1}=9.244$ TeV& $m_{h^\pm_2}=31.96$ TeV&$m_{h_3^\pm}=32.87$ TeV &\multicolumn{1}{c}{}\\
\cline{1-3}
\end{tabular}
\caption{Spectrum of the scalar sector in the triplet/singlet extension of the MSSM consistent with the SM-like Higgs with $\sim125$ GeV mass.}\label{spectr}
\end{center}
\end{table}

Finally let us consider the simplest case where we have superfields in the singlet, doublet and triplet representation of $SU(2)_w$. The model contains a singlet superfield and a triplet superfield with $Y=0$ on the top of the MSSM superfield content \cite{Basak,PBAC1}. The superpotential is 
\begin{align}\label{Wtnssm}
\mathcal{\hat W}_{TNMSSM}=&\,y_u \hat U \hat H_u \hat Q - y_d \hat D \hat H_d \hat Q - y_e  \hat E \hat H_d \hat L \\
&\,+ \lambda_S \hat S \hat H_u \hat H_d + \lambda_{TS}\hat S \hat T^2 + \lambda_T \hat H_u\hat T \hat H_d +\frac{\kappa}{3} \hat S^3\nn
\end{align}
It contains no mass parameters and the potential, in the limit of small trilinear terms $A_i\to0$, exhibit a global $U(1)$ symmetry which can be softly broken giving rise to a pseudo-Nambu-Goldstone mode in the form of a light pseudoscalar \cite{PBAC2}. The RG equations for the gauge couplings at one-loop are the same of the TMSSM, Eq. \ref{tmssmrg}. As we have shown in the previous sections, when the RG equations for the gauge couplings are evaluated at two-loop all the dimensionless couplings contribute to the running and hence the couplings appearing in Eq. \ref{Wtnssm} can drive the gauge couplings unification.

In Figure \ref{tnssmuniH} (a) we present the scenario where we have the gauge couplings unification with RG equations calculated at two-loop. We have not presented the running of the gauge couplings at one-loop because there is no unification in this case, similarly to the TMSSM case. In Figure \ref{tnssmuniH} (b) we plot the running of the dimensionless couplings at one-loop (dashed lines) and at two-loop (solid lines). Again at one-loop there is the appearance of a Landau pole in the evolution of the couplings whereas it is not present with RG equations calculated at two-loop. Even in this case the couplings are in the perturbative range until the Planck scale.

In this case we have requested that the values of the couplings at the electroweak scale are compatible with a Higgs boson with mass $\sim125$ GeV. In Table \ref{spectr} we give the spectrum of the neutral and charged scalars at the electroweak scale, showing that apart from the $\sim125$ GeV Higgs, the other states are essentially decoupled. The lightest scalar Higgs is more than $96\%$ up-type whereas the lightest pseudoscalar is more than $99\%$ singlet-type. This means that $h_1$ plays the role of the SM-like scalr Higgs boson whereas the presence of a light pseudoscalar which is dominated by the singlet does not affect the constraint coming from $B$-observables \cite{PBAC3}.

The complete analysis of the stability of the potential is beyond the purpose of this work. Here we shall confine ourselves to point out the possibility of a SM-like Higgs boson in case of $\lambda$SUSY when we demand for the gauge couplings unification. Concerning the Higgs mass, in theories where the dimensionless couplings of the superpotential are large ($\vec\lambda\gsim1$), the one-loop correction to the Higgs mass are less important respect to the usual case $\vec\lambda<1$ \cite{staublambdasusy}.

\section{$\vec\lambda$ at the electroweak and at the Planck scale}\label{lamWH}

\begin{figure}[t]
\centering
\mbox{
\hspace{-1.75cm}
\subfigure[]{
\includegraphics[width=0.4\textwidth]{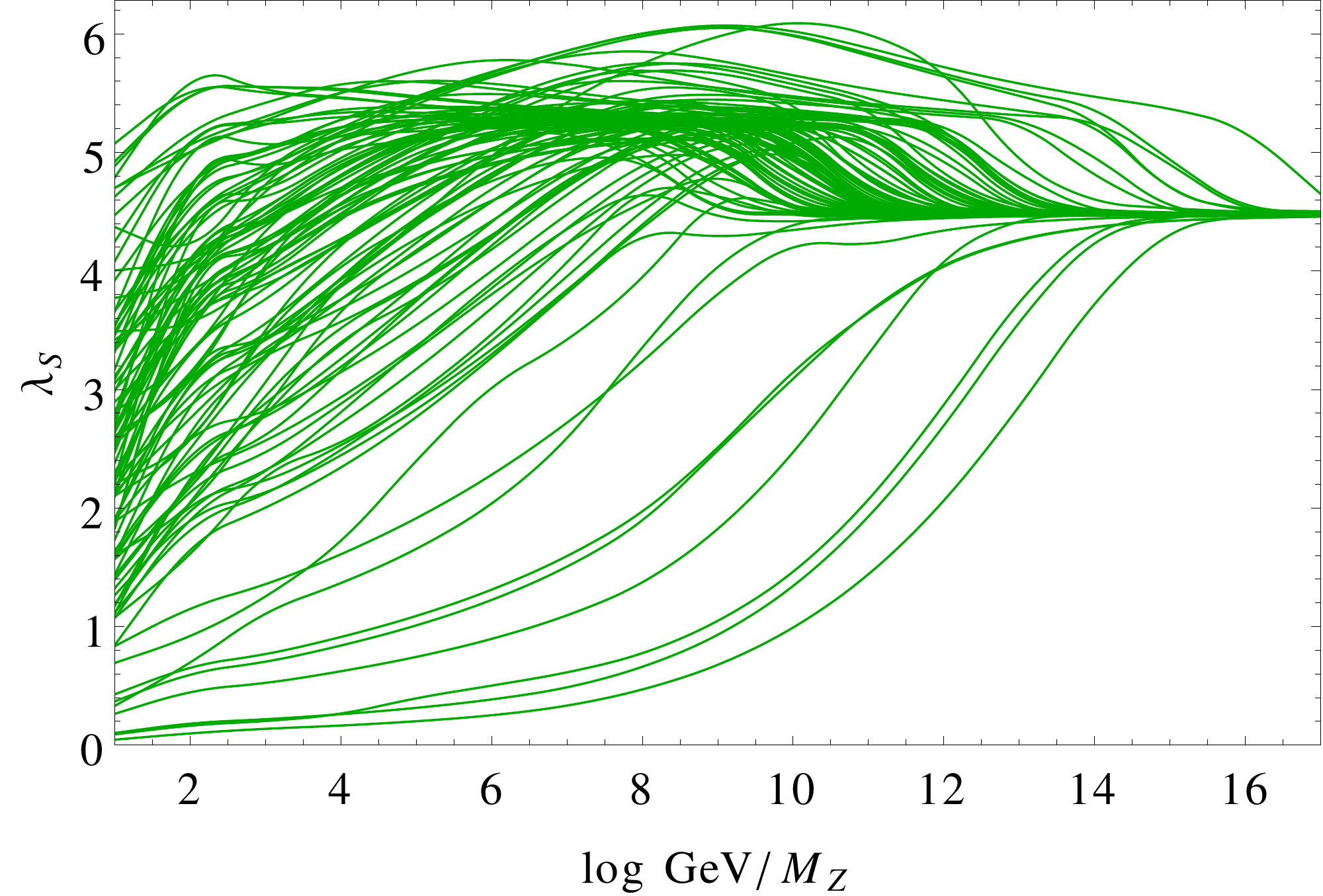}}
\subfigure[]{\includegraphics[width=0.4\textwidth]{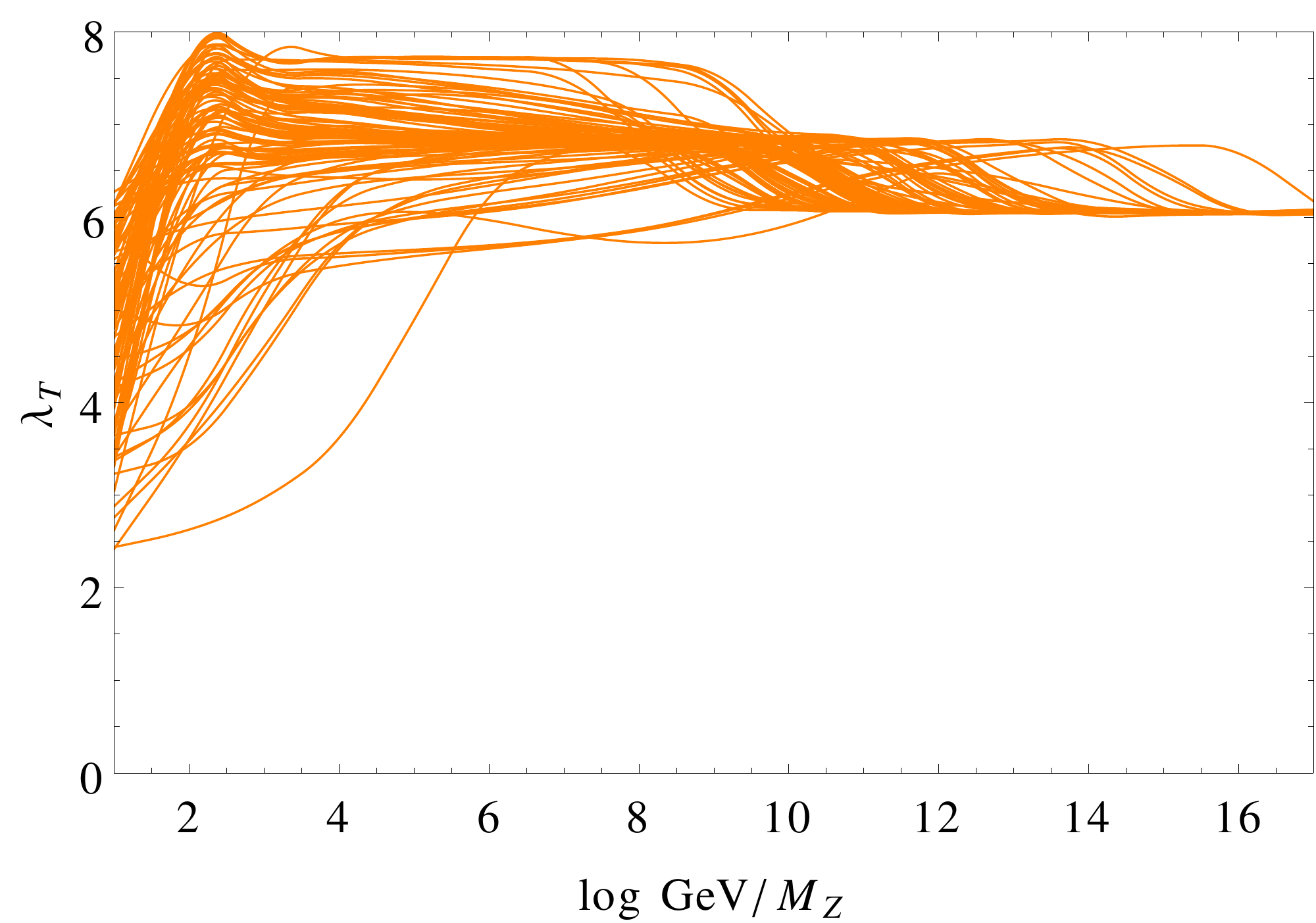}}
\subfigure[]{\includegraphics[width=0.4\textwidth]{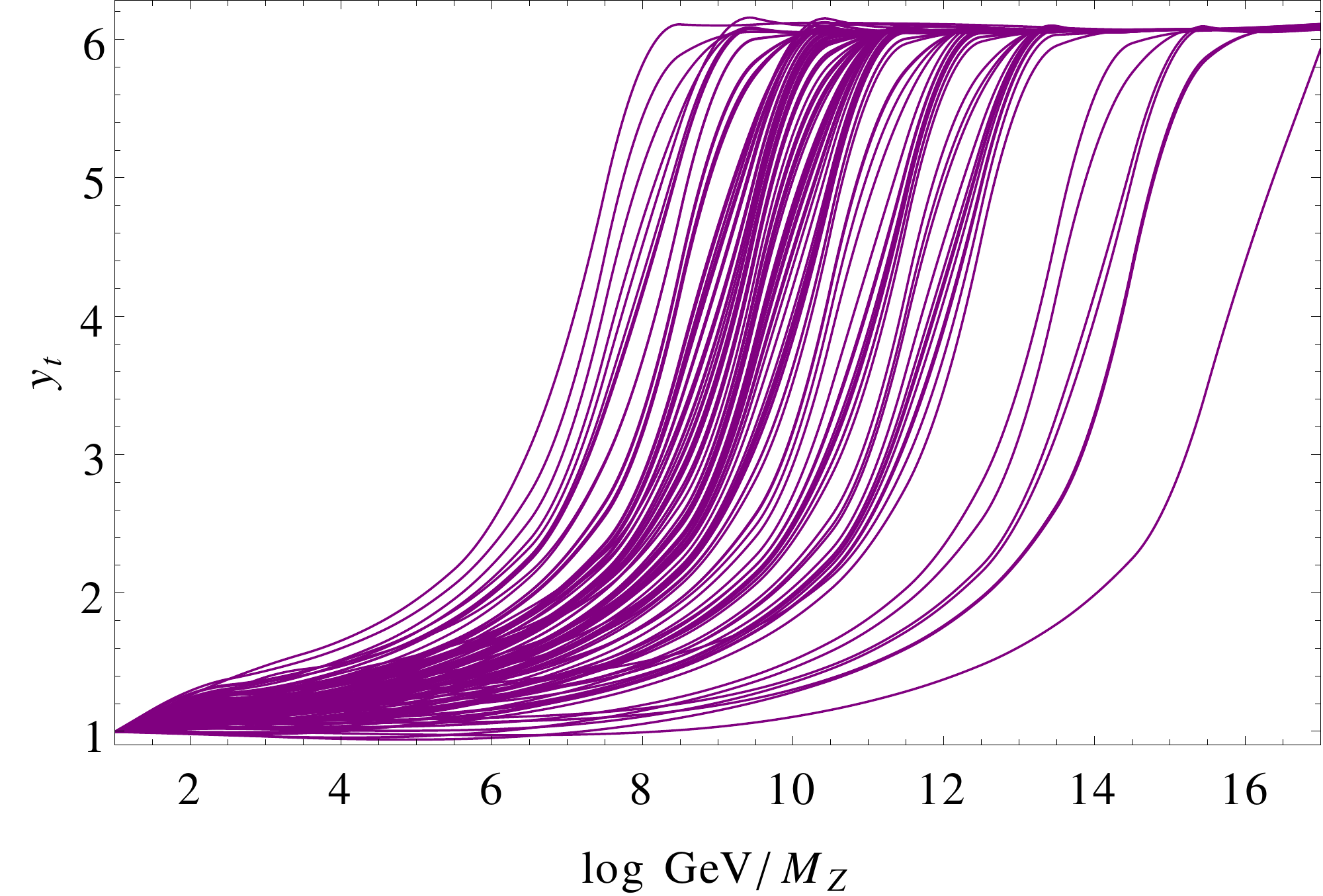}}}
\mbox{\subfigure[]{
\includegraphics[width=0.4\textwidth]{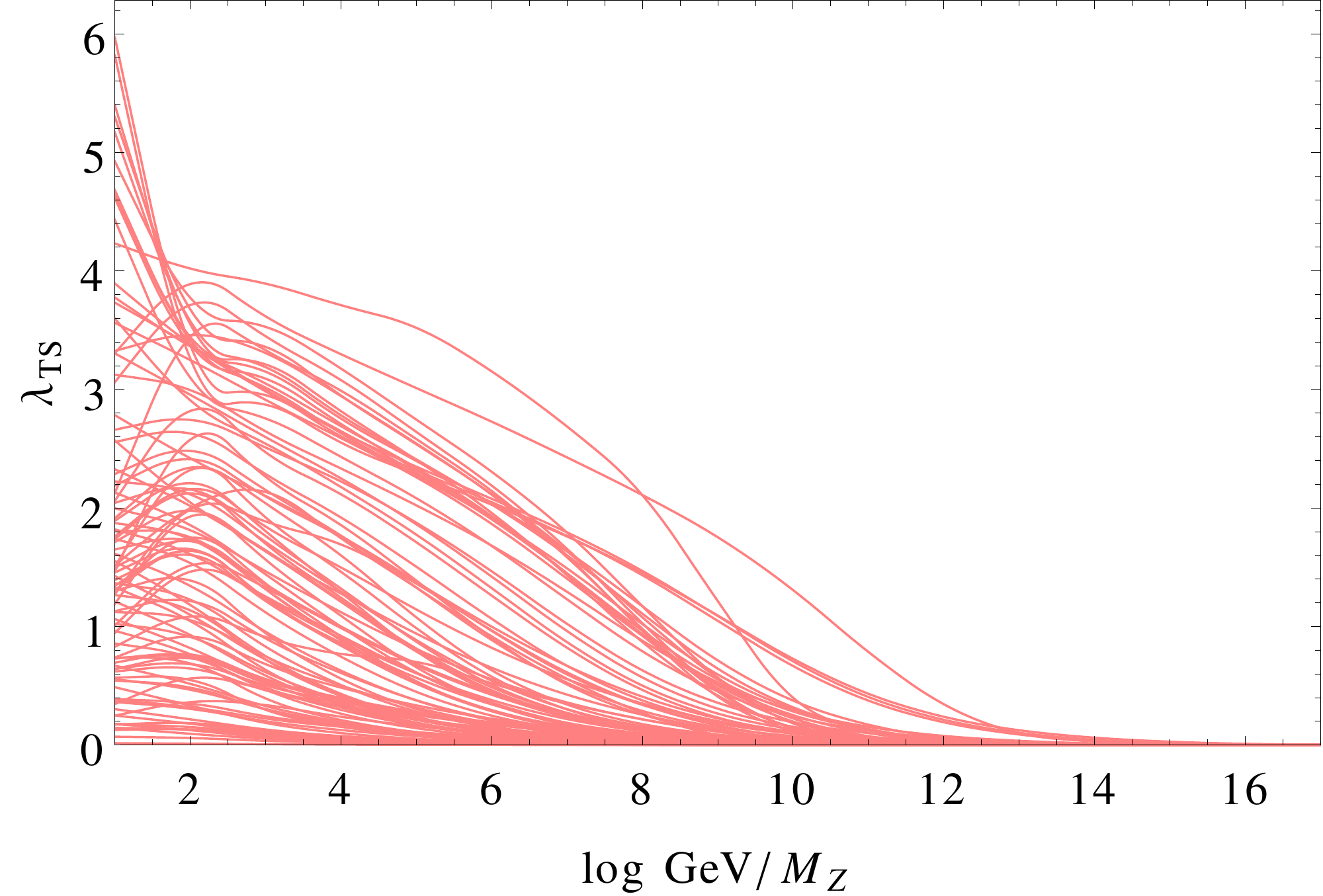}}
\subfigure[]{\includegraphics[width=0.4\textwidth]{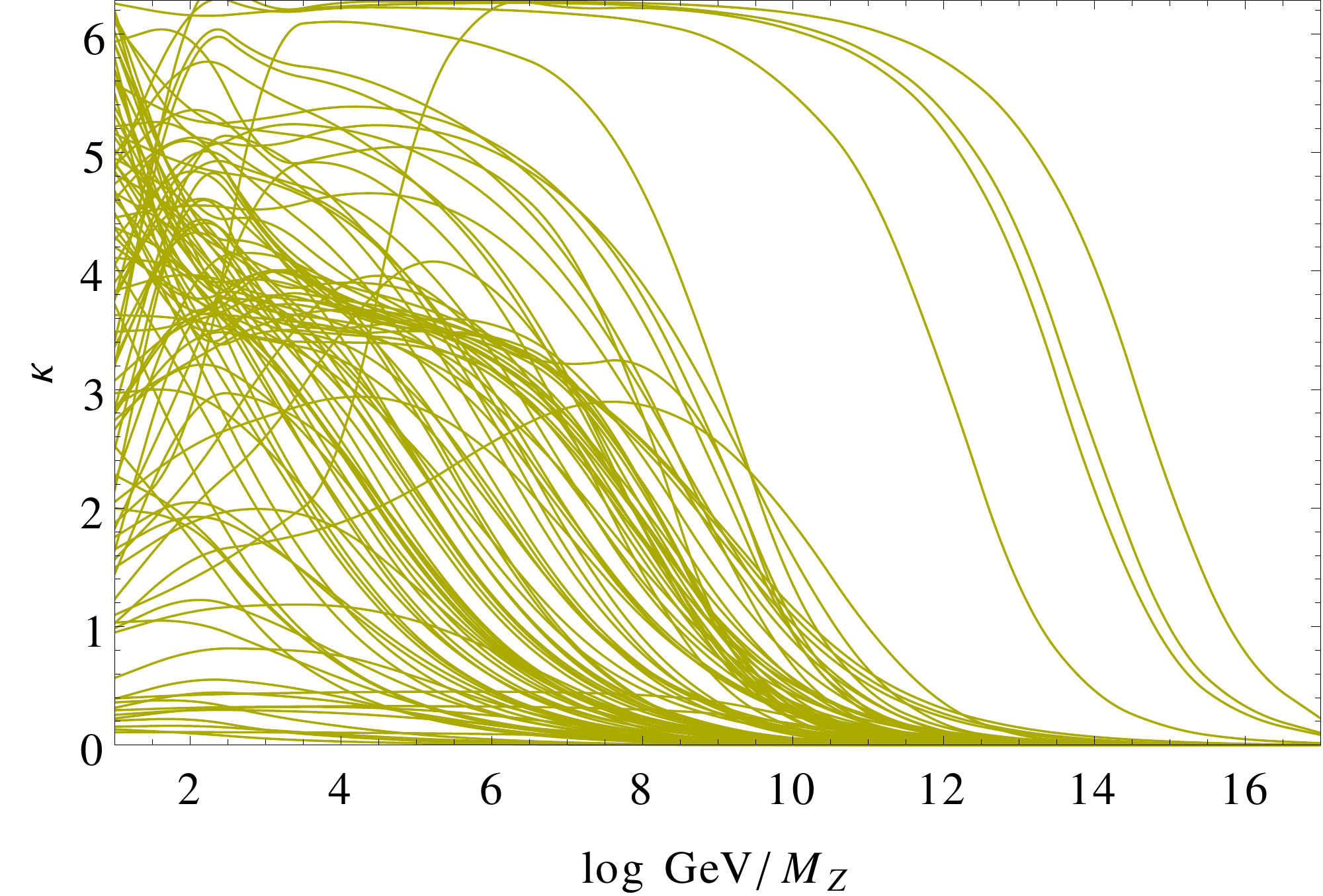}}}
\caption{Running values of $\lambda_S$ (a), $\lambda_T$ (b), $y_t$ (c), $\lambda_{TS}$ (d) and $\kappa$ (e). For each point ($\lambda_S$, $\lambda_T$, $y_t$, $\lambda_{TS}$, $\kappa$) at the electroweak scale, the gauge couplings unification is achieved at two-loop.}
 \label{fixedpWH}
\end{figure}

In the previous sections we have analyzed various supersymmetric theories, from the minimal formulation to extensions including superfield(s) in higher representations of $SU(2)_w$. We have discussed the possibility of the gauge couplings unification at one- and two-loop. We have shown that the unification can be achieved at two-loop order in theories with triplet superfield(s) if $\vec\lambda\gsim1$. Moreover the unification of the gauge couplings in this class of models seems to be related to the appearance of UV fixed points for the dimensionless couplings. This is drastically different from the evolution of the $\vec\lambda$'s at one-loop which usually present a Landau pole at intermediate scale.
 
Looking at Figure \ref{tnssmuniH} (b) we can see that $\lambda_S$, $\lambda_T$ and $y_t$ evolve to a finite value near the Planck scale but $\kappa$ and $\lambda_{TS}$ tend to be zero at high energies. The value of the couplings around the Planck scale is crucial in the determination of the trivial or nontrivial nature of the UV fixed point. It is then interesting to test if the scenario of Figure \ref{tnssmuniH} (b) is a coincidence or not. The result is presented in Figure \ref{fixedpWH}.

In Figure \ref{fixedpWH} (a)-(e) we have compared the values of $\lambda_S$, $\lambda_T$, $y_t$, $\lambda_{TS}$ and $\kappa$ from the electroweak to the Planck scale. We have considered a sample of 100 different values, in the range $\lambda_i,\,\kappa\in[0,\,2\pi]$, of the dimensionless couplings at the electroweak scale which realize the gauge couplings unification at two-loop. Here however we haven't requested the presence of a $\sim125$ GeV Higgs boson in the spectrum. Nevertheless the evolution of the couplings is very interesting. Regardless of their values at the electroweak scale, the dimensionless parameters of the superpotential approach a fixed value at the Planck scale. In particular, $\lambda_{TS}$ and $\kappa$ tend to be zero at high energies whereas $\lambda_S$ and $\lambda_T$ and $y_t$ have a finite value around the Planck scale. It is interesting to note that the electroweak value of $\lambda_S$, $\lambda_{TS}$ and $\kappa$ can span the whole range considered whereas to achieve the gauge couplings unification we need to have $\lambda_T>2$. However it is clear that the scenario of Figure \ref{tnssmuniH} (b) does not present a particular situation. Conversely the nontrivial nature of the UV fixed point appears to be a general feature of the triplet/singlet extension of the MSSM if we insist on the gauge couplings unification.

\section{Conclusions}\label{concl}

The aim of this paper was to discuss the impact of two-loop RG equations on the running of the gauge couplings of $SU(3)_c\times SU(2)_w\times U(1)_Y$. The MSSM is the simplest model featuring the gauge couplings unification and in this case the inclusion of the two-loop RG equation does not considerably affect the running of the gauge couplings. This is due to the fact that in the MSSM beside the gauge couplings the only other relevant dimensionless parameter is the Yukawa coupling of the top quark. In the simplest extension of the MSSM, \textit{i.e.} the NMSSM, the situation is similar. Although there are two dimensionless parameters in the superpotential, $\lambda$ and $\kappa$, the running of the gauge couplings at two-loop is affected only by $\lambda$. This is because the coupling $\kappa$ is the self-interaction of the singlet and thus cannot affect directly the RG equations for the couplings of $SU(3)_c\times SU(2)_w\times U(1)_Y$. 

If we consider the extensions of the MSSM with superfields in higher representations of $SU(2)_w$ the situation is quite different. The gauge couplings do not unify at any scale at one-loop. The main reason is the $SU(2)_w$ charge of the triplet which affect the $c_{g_2}$ coefficient of the $\beta$-function. At two-loop, however, the running of the dimensionless couplings is governed by a system of coupled differential equations. The evolution of the gauge couplings are then affected by the other dimensionless parameters. In this case the gauge couplings unification at two-loop can be achieved if the electroweak value of the superpotential's parameters is large enough. For such large values of $\vec\lambda$ the RG equations develop a Landau pole at one-loop below the Planck scale. The Landau pole is not present in the evolution of the couplings at two-loop. Conversely $\vec\lambda$ approach a UV fixed point at very-high energies. The asymptotic vanishing of the $\beta$-functions of $\vec\lambda$ is a consequence of the competition between the one- and two-loop contribution, which is possible in $\lambda$SUSY models. 

In the case of TNMSSM, where a singlet and a triplet superfields with $Y=0$ are added to the superfield content of the MSSM, we have presented a scenario where the gauge couplings unify around the Planck scale and the electroweak values of the dimensionless parameters are compatible with a Higgs boson with $\sim125$ GeV of mass, with the rest of the spectrum essentially decoupled. The nontrivial nature of the UV fixed point appears to be a general feature of the model if we request the gauge couplings unification, because the value of $\lambda_S$, $\lambda_T$, $\lambda_{TS}$, $\kappa$ and $y_t$ around the Planck scale seems to be roughly independent from their value at the electroweak scale.

\appendix

\small

\section{Two-Loop Expressions of the RG Equations}\label{APPtl}
We present the explicit expression of the RG equations for the dimensionless couplings at one- and two-loop for the various models considered. Regarding the Yukawas, we include only the Yukawa of the top quark because it is the dominant one. We write in general
\bea
\frac{d\,g_i}{d t}=\beta^{(1)}_{g_i} + \beta^{(2)}_{g_i}
\eea 
where $\beta^{(1)}_{g_i}$ and $\beta^{(2)}_{g_i}$ are the one-loop and the two-loop parts respectively. Similarly, for each dimensionless coupling appearing in the superpotential we have
\bea
\frac{d\,\lambda_i}{d t}=\beta^{(1)}_{\lambda_i} + \beta^{(2)}_{\lambda_i}
\eea 
where $\lambda_i$ denotes a generic dimensionless coupling.

\subsection{MSSM}
In the MSSM there are no dimensionless parameters in the superpotential but the Yukawa of the top quark. The one-loop contribution to the $\beta$-functions for the gauge couplings are already given in Eq. \ref{mssmrg}.

\begin{flalign}
\beta^{(2)}_{g_1}&=\frac{g_1^3}{6400\pi^4}\left(199 g_1^2+135 g_2^2+440 g_3^2-130 y_t^2\right)&&\\
\beta^{(2)}_{g_2}&=\frac{g_2^3}{1280\pi^4}\left(9 g_1^2+125 g_2^2+120 g_3^2-30 y_t^2\right)\\
\beta^{(2)}_{g_3}&=\frac{g_3^3}{6400\pi^4}\left(11 g_1^2+45 g_2^2+70 g_3^2-20 y_t^2\right)\\
\beta^{(1)}_{y_t}&=\frac{y_t}{16\pi^2}\left(3y_t^2-\left(\frac{13}{15}g_1^2+3g_2^2+\frac{16}{3}g_3^2-3y_t^2\right)\right)\\
\beta^{(2)}_{y_t}&=\frac{1}{256\pi^4}\left(\frac{2}{5}g_1^2 y_t^3+6g_2^2 y_t^3-13 y_t^5+y_t\left(\frac{2743}{450}g_1^4+g_1^2 g_2^2+\frac{15}{2}g_2^4+\frac{136}{45}g_1^2 g_3^2\right.\right.\nn\\
&\left.\left.\qquad\qquad\quad+8g_2^2 g_3^2-\frac{16}{9}g_3^4+\frac{4}{5}(g_1^2+20g_3^2)y_t^2-9y_t^4\right)\right)
\end{flalign}

\subsection{NMSSM}
The addition of the singlet superfield $\hat S$ to the MSSM introduce two dimensionless couplings $\lambda$ and $\kappa$, \textit{cf.} Eq. \ref{WNMSSM}.

\begin{flalign}
\beta^{(2)}_{g_1}&=\frac{g_1^3}{6400\pi^4}\left(199 g_1^2+135 g_2^2+440 g_3^2-30\lambda^2-130 y_t^2\right)&&\\
\beta^{(2)}_{g_2}&=\frac{g_2^3}{1280\pi^4}\left(9 g_1^2+125 g_2^2+120 g_3^2-10\lambda^2-30 y_t^2\right)\\
\beta^{(2)}_{g_3}&=\frac{g_3^3}{6400\pi^4}\left(11 g_1^2+45 g_2^2+70 g_3^2-20 y_t^2\right)\\
\beta^{(1)}_{y_t}&=\frac{y_t}{16\pi^2}\left(3y_t^2-\left(\frac{13}{15}g_1^2+3g_2^2+\frac{16}{3}g_3^2-\lambda^2-3y_t^2\right)\right)
\end{flalign}
\begin{flalign}
\beta^{(2)}_{y_t}&=\frac{1}{256\pi^4}\left(\frac{2}{5}g_1^2 y_t^3+6g_2^2 y_t^3-13 y_t^5-3\lambda^2 y_t^3+y_t\left(\frac{2743}{450}g_1^4+g_1^2 g_2^2+\frac{15}{2}g_2^4+\frac{136}{45}g_1^2 g_3^2\right.\right.\nn\\
&\left.\left.\qquad\qquad\,\,\,+8g_2^2 g_3^2-\frac{16}{9}g_3^4+\frac{4}{5}(g_1^2+20g_3^2)y_t^2-9y_t^4-2\kappa^2\lambda^2-3\lambda^4\right)\right)\\
\beta^{(1)}_\lambda&=\frac{\lambda}{16\pi^2}\left(4\lambda^2+2\kappa^2+3y_t^2-\frac{3}{5}g_1^2-3g_2^2\right)\\
\beta^{(2)}_\lambda&=\frac{-\lambda}{12800\pi^4}\Big(500\lambda^4+600\kappa^2\lambda^2+400\kappa^4-207g_1^4-90 g_1^2 g_2^2-375g_2^4-40g_1^2 y_t^2-800g_3^2 y_t^2\nn\\
&\qquad\qquad\quad+450 y_t^4-30\lambda^2\left(2g_1^2+10g_2^2-15y_t^2\right)\Big)\\
\beta^{(1)}_\kappa&=\frac{3\kappa}{8\pi^2}\left(\kappa^2+\lambda^2\right)\\
\beta^{(2)}_\kappa&=\frac{-3\kappa}{640\pi^4}\Big(20\kappa^4+20\kappa^2\lambda^2+\lambda^2(10\lambda^2+15y_t^2-3g_1^2-15g_2^2)\Big)
\end{flalign}

\subsection{TMSSM}
In the TMSSM there is an extra triplet superfield with $Y=0$ on top of the MSSM superfield content. This results in one extra dimensionless parameter in the superpotential, which we call $\lambda$, \textit{cf.} Eq. \ref{WTMSSM}. The one-loop part of the $\beta$-functions for the gauge couplings are given in Eq. \ref{tmssmrg}.

\begin{flalign}
\beta^{(2)}_{g_1}&=\frac{g_1^3}{6400\pi^4}\left(199 g_1^2+135 g_2^2+440 g_3^2-45\lambda^2-130 y_t^2\right)&&\\
\beta^{(2)}_{g_2}&=\frac{g_2^3}{1280\pi^4}\left(9 g_1^2+125 g_2^2+120 g_3^2-35\lambda^2-30 y_t^2\right)\\
\beta^{(2)}_{g_3}&=\frac{g_3^3}{6400\pi^4}\left(11 g_1^2+45 g_2^2+70 g_3^2-20 y_t^2\right)\\
\beta^{(1)}_{y_t}&=\frac{y_t}{16\pi^2}\left(3y_t^2-\left(\frac{13}{15}g_1^2+3g_2^2+\frac{16}{3}g_3^2-3y_t^2-\frac{3}{2}\lambda^2\right)\right)\\
\beta^{(2)}_{y_t}&=\frac{1}{256\pi^4}\left(\frac{2}{5}g_1^2 y_t^3+6g_2^2 y_t^3-13 y_t^5-\frac{9}{2}\lambda^2 y_t^3+y_t\left(\frac{2743}{450}g_1^4+g_1^2 g_2^2+\frac{15}{2}g_2^4+\frac{136}{45}g_1^2 g_3^2\right.\right.\nn\\
&\left.\left.\qquad\qquad\quad+8g_2^2 g_3^2-\frac{16}{9}g_3^4+6g_2^2 \lambda^2-\frac{15}{4}\lambda^4+\frac{4}{5}(g_1^2+20g_3^2)y_t^2-9y_t^4\right)\right)\\
\beta^{(1)}_\lambda&=\frac{\lambda}{16\pi^2}\left(4\lambda^2+3y_t^2-\frac{3}{5}g_1^2-7g_2^2\right)\\
\beta^{(2)}_\lambda&=\frac{-\lambda}{12800\pi^4}\Big(525\lambda^4-207g_1^4-90 g_1^2 g_2^2-2075g_2^4-40g_1^2 y_t^2-800g_3^2 y_t^2+450y_t^4\nn\\
&\qquad\qquad\quad-5\lambda^2(6g_1^2+110g_2^2-75y_t^2)\Big)
\end{flalign}

\subsection{TcMSSM}
In the case of TcMSSM there are four extra dimensionless couplings w.r.t. the MSSM case. We have called them $\lambda$, $\zeta$, $\xi_u$ and $\xi_d$. The one-loop part of the $\beta$-functions for the gauge couplings are given in Eq. \ref{tcmssmrg}.   

\begin{flalign}
\beta^{(2)}_{g_1}&=\frac{g_1^3}{1280\pi^4}\left(83 g_1^2+171 g_2^2+88 g_3^2-36\zeta^2-9\lambda^2-45\xi_u^2-45\xi_d^2-26 y_t^2\right)&&\\
\beta^{(2)}_{g_2}&=\frac{g_2^3}{1280\pi^4}\left(57 g_1^2+485 g_2^2+120 g_3^2-60\zeta^2-35\lambda^2-70\xi_u^2-70\xi_d^2-30 y_t^2\right)\\
\beta^{(2)}_{g_3}&=\frac{g_3^3}{6400\pi^4}\left(11 g_1^2+45 g_2^2+70 g_3^2-20 y_t^2\right)\\
\beta^{(1)}_{y_t}&=\frac{y_t}{16\pi^2}\left(3y_t^2-\left(\frac{13}{15}g_1^2+3g_2^2+\frac{16}{3}g_3^2-3y_t^2-\frac{3}{2}\lambda^2-6\xi_u^2\right)\right)\\
\beta^{(2)}_{y_t}&=\frac{1}{256\pi^4}\Bigg(\frac{2}{5}g_1^2 y_t^3+6g_2^2 y_t^3-13 y_t^5-\frac{9}{2}\lambda^2y_t^3-18\xi_u^2 y_t^3+\frac{y_t}{900}\big(1350\lambda^2(4g_2^2-\zeta^2-6\xi_u^2-6\xi_d^2)\nn\\
&\qquad\qquad-3375\lambda^4+2(4147g_1^4+450 g_1^2 g_2^2+11475 g_2^2+1360 g_1^2 g_3^2+3600 g_2^2 g_3^2-800 g_3^4\nn\\
&\qquad\qquad-21600\xi_u^4+360y_t^2(g_1^2+20g_3^2)-4050y_t^4+540\xi_u^2(6g_1^2+20g_2^2-5\zeta^2-15y_t^2))\big)\Bigg)\\
\beta^{(1)}_\lambda&=\frac{\lambda}{16\pi^2}\left(4\lambda^2+6\xi_u+6\xi_d+\zeta^2-\frac{3}{5}g_1^2-7g_2^2\right)\\
\beta^{(2)}_\lambda&=\frac{-\lambda}{2560\pi^4}\Bigg(105\lambda^4+20\zeta^4-63g_1^4-18g_1^2 g_2^2-695g_2^4-72g_1^2\xi_d^2-240g_2^2\xi_d^2+480\xi_d^4\nn\\
&\qquad\qquad\quad-72g_1^2\xi_u^2-240g_2^2\xi_u^2+480\xi_u^4+2\zeta^2\big(40\xi_d^2+40\xi_u^2+15\lambda^2-12g_1^2-20g_2^2-8g_1^2 y_t^2\big)\nn\\
&\qquad\qquad\quad-160g_3^2 y_t^2+180\xi_u^2 y_t^2+90y_t^4+\lambda^2\big(240\xi_d^2+240\xi_u^2+75y_t^2-6g_1^2-110g_2^2\big)\Bigg)\\
\beta^{(1)}_\zeta&=\frac{\zeta}{16\pi^2}\left(3\zeta^2+\lambda^2+\frac{2}{5}\left(5\xi_d^2+5\xi_u^2-6(g_1^2+5g_2^2)\right)\right)\\
\beta^{(2)}_\zeta&=\frac{-\zeta}{6400\pi^4}\Bigg(150\zeta^4+75\lambda^4-648g_1^4-480g_1^2 g_2^2-3300g_2^4+30g_1^2\xi_d^2+50g_2^2\xi_d^2+600\xi_d^4\nn\\
&\qquad\qquad\quad+30g_1^2\xi_u^2+50g_2^2\xi_u^2+600\xi_u^4-10\zeta^2\big(6g_1^2+30g_2^2-5\lambda^2-10\xi_d^2-10\xi_u^2\big)\nn\\
&\qquad\qquad\quad+300\xi_u^2 y_t^2+5\lambda^2\big(60\xi_d^2+60\xi_u^2+15y_t^2+5g_2^2-3g_1^2\big)\Bigg)\\
\beta^{(1)}_{\xi_u}&=\frac{\xi_u}{16\pi^2}\left(14\xi_u^2+6y_t^2+\zeta^2+3\lambda^2-\frac{9}{5}g_1^2-7g_2^2\right)\\
\beta^{(2)}_{\xi_u}&=\frac{-\xi_u}{12800\pi^4}\Bigg(100\zeta^4+375\lambda^4-999g_1^4-570g_1^2 g_2^2-3475 g_2^4-660g_1^2\xi_u^2-2300g_2^2\xi_u^2-6000\xi_u^4\nn\\
&\qquad\qquad\quad-150\lambda^2\big(4g_2^2-6\xi_d^2-8\xi_u^2\big)-100\zeta^2\big(2g_2^2-2\lambda^2-\xi_d^2-6\xi_u^2\big)-80g_1^2 y_t^2-1600g_3^2 y_t^2\nn\\
&\qquad\qquad\quad+2400\xi_u^2 y_t^2+900 y_t^4\Bigg)\\
\beta^{(i)}_{\xi_d}&=\beta^{(i)}_{\xi_u}\Big|_{\xi_u\to\xi_d,\,y_t\to0}
\end{flalign}

\subsection{Triplet/Singlet Extended MSSM}

In this model the MSSM superfield content is extended with two extra superfields, a singlet and triplet with $Y=0$. Hence in the superpotential there are four extra dimensionless couplings w.r.t. the MSSM, which we have called $\lambda_S$, $\lambda_T$, $\lambda_{TS}$ and $\kappa$. The one-loop contribution to the $\beta$-functions of the gauge couplings is the same of the TMSSM and it is given in Eq. \ref{tmssmrg}.

\begin{flalign}
\beta^{(2)}_{g_1}&=\frac{g_1^3}{6400\pi^4}\left(199 g_1^2+135 g_2^2+440 g_3^2-45\lambda_T^2-30\lambda_S^2-130 y_t^2\right)&&\\
\beta^{(2)}_{g_2}&=\frac{g_2^3}{1280\pi^4}\left(9 g_1^2+245 g_2^2+120 g_3^2-35\lambda_T^2-80\lambda_{TS}^2-10\lambda_S^2-30 y_t^2\right)\\
\beta^{(2)}_{g_3}&=\frac{g_3^3}{6400\pi^4}\left(11 g_1^2+45 g_2^2+70 g_3^2-20 y_t^2\right)\\
\beta^{(1)}_{y_t}&=\frac{y_t}{16\pi^2}\left(3y_t^2-\left(\frac{13}{15}g_1^2+3g_2^2+\frac{16}{3}g_3^2-3y_t^2+\frac{3}{2}\lambda_T^2+\lambda_S^2\right)\right)\\
\beta^{(2)}_{y_t}&=\frac{1}{256\pi^4}\left(\frac{2}{5}g_1^2 y_t^3+6g_2^2 y_t^3-\frac{9}{2}\lambda_T^2 y_t^3-3\lambda_S^2 y_t^3-13 y_t^5+y_t\left(\frac{2743}{450}g_1^4+g_1^2 g_2^2+\frac{27}{2}g_2^4\right.\right.\nn\\
&\left.\left.\qquad\qquad\quad+\frac{136}{45}g_1^2 g_3^2+8g_2^2 g_3^2-\frac{16}{9}g_3^4-\frac{15}{4}\lambda_T^4-6\lambda_{TS}^2\lambda_S^2-2\kappa^2\lambda_S^2-3\lambda_S^4\right.\right.\nn\\
&\left.\left.\qquad\qquad\quad+\frac{4}{5}(g_1^2+20g_3^2)y_t^2-9y_t^4+\frac{3}{2}\lambda_T^2(4g_2^2-4\lambda_{TS}^2-2\lambda_S^2)\right)\right)\\
\beta^{(1)}_{\lambda_S}&=\frac{\lambda_S}{16\pi^2}\left(4\lambda_S^2+3\lambda_T^2+6\lambda_{TS}^2+2\kappa^2+3y_t^2-\frac{3}{5}g_1^2-3g_2^2\right)&&\\
\beta^{(2)}_{\lambda_S}&=\frac{-\lambda_S}{12800\pi^4}\Bigg(375\lambda_T^4+2400\lambda_{TS}^4+400\kappa^4+500\lambda_S^4-207g_1^4-90g_1^2 g_2^2-675g_2^4-60g_1^2\lambda_S^2\nn\\
&\qquad\qquad\quad-300g_2^2\lambda_S^2+600\kappa^2\lambda_S^2-600\lambda_{TS}^2\big(4g_2^2-2\kappa^2-\lambda_S^2\big)-40g_1^2y_t^2-800g_3^2 y_t^2\nn\\
&\qquad\qquad\quad+450\lambda_S^2 y_t^2+450y_t^4-75\lambda_T^2\big(8g_2^2-16\lambda_{TS}^2-8\lambda_S^2-3y_t^2\big)\Bigg)\\
\beta^{(1)}_{\lambda_T}&=\frac{\lambda_T}{16\pi^2}\left(4\lambda_T^2+4\lambda_{TS}^2+2\lambda_{S}^2+3y_t^2-\frac{3}{5}g_1^2-7g_2^2\right)\\
\beta^{(2)}_{\lambda_T}&=\frac{-\lambda_T}{12800\pi^4}\Bigg(525\lambda_T^4+2000\lambda_{TS}^4+300\lambda_S^4+200\kappa^2\lambda_S^2-207g_1^4-90g_1^2 g_2^2-2075g_2^4\nn\\
&\qquad\qquad\quad+200\lambda_{TS}^2\big(2\kappa^2+5\lambda_S^2\big)-40g_1^2 y_t^2-800 g_3^2 y_t^2+150\lambda_S^2 y_t^2+450y_t^4\nn\\
&\qquad\qquad\quad-5\lambda_T^2\big(6g_1^2+110g_2^2-160\lambda_{TS}^2-80\lambda_S^2-75y_t^2\big)\Bigg)\\
\beta^{(1)}_{\lambda_{TS}}&=\frac{\lambda_{TS}}{8\pi^2}\Big(\lambda_T^2+\kappa^2+\lambda_S^2+7\lambda_{TS}^2-4g_2^2\Big)
\end{flalign}
\begin{flalign}
\beta^{(2)}_{\lambda_{TS}}&=\frac{-\lambda_{TS}}{640\pi^4}\Bigg(320\lambda_{TS}^4+15\lambda_T^4+20\kappa^4+10\lambda_S^4-140g_2^4-3g_1^2\lambda_S^2-15g_2^2\lambda_S^2+20\kappa^2\lambda_S^2\nn\\
&\qquad\qquad-20\lambda_{TS}^2\big(6g_2^2-5\kappa^2-2\lambda_S^2\big)+15\lambda_S^2 y_t^2+\lambda_T^2\big(50\lambda_{TS}^2+25\lambda_S^2+15 y_t^2-3g_1^2+5g_2^2\big)\Bigg)\\
\beta^{(1)}_\kappa&=\frac{3\kappa}{8\pi^2}\big(\kappa^2+\lambda_S^2+\lambda_{TS}^2\big)\\
\beta^{(2)}_\kappa&=\frac{-3\kappa}{640\pi^4}\Bigg(120\lambda_{TS}^4+20\kappa^4-30\lambda_{TS}^2\big(4g_2^2-\lambda_T^2-2\kappa^2\big)+20\kappa^2\lambda_S^2+\lambda_S^2\big(15\lambda_T^2\nn\\
&\qquad\qquad\quad+10\lambda_S^2+15 y_t^2-3g_1^2-15g_2^2\big)\Bigg)
\end{flalign}

\end{document}